\documentclass[12pt]{article}
\usepackage{times}
\usepackage{epsfig}
\usepackage{color}
\usepackage{axodraw}
\def\beq{\begin{equation}}
\def\eeq{\end{equation}}
\def\bea{\begin{eqnarray}}
\def\eea{\end{eqnarray}}

\def\nn{\nonumber}
\def\Eq#1{Eq.~(\ref{#1})}

\begin{document}
\begin{titlepage}

\renewcommand{\thefootnote}{\fnsymbol{footnote}}
\begin{flushright}
    LPN11-52, TTP11-26, IFIC/11-45  \\ 
\end{flushright}
\par \vspace{10mm}

\begin{center}

{\Large \bf Charge asymmetries of top quarks at hadron colliders revisited}

\vspace{8mm}

{\bf Johann H. K\"uhn$^a$~\footnote{E-mail: johann.kuehn@kit.edu}} and
{\bf Germ\'an Rodrigo$^{a,b}$~\footnote{E-mail: german.rodrigo@csic.es}}

\vspace{5mm}
$~^a$ Institut f\"ur Theoretische Teilchenphysik,
Karlsruher Institut f\"ur Technologie, \\
D-76133, Karlsruhe, Germany. \\
\vspace{5mm}
$~^b$ Instituto de F\'{\i}sica Corpuscular, \\
Consejo Superior de Investigaciones Cient\'ificas-Universitat de Val\`encia, \\
Parc Cient\'ific, E-46980 Paterna (Valencia), Spain. \\

\vspace{5mm}
\end{center}

\par \vspace{2mm}
\begin{center} {\large \bf Abstract} \end{center}

\begin{quote}
A sizeable difference in the differential production cross section of
top- compared to antitop-quark production, denoted charge asymmetry, has
been observed at the Tevatron. The experimental results seem to
exceed the theory predictions based on the Standard Model by a
significant amount and have triggered a large number of suggestions for
''new physics''. In the present paper the Standard Model predictions for
Tevatron and LHC experiments are revisited. This includes a reanalysis
of electromagnetic as well as weak corrections, leading to a shift of
the asymmetry by roughly a factor 1.1 when compared to the results of
the first papers on this subject. The impact of cuts on the 
transverse momentum of the top-antitop system is studied. Restricting
the $t\bar t$ system to a transverse momentum less than 20~GeV leads to
an enhancement of the asymmetries by factors between 1.3 and 1.5,
indicating the importance of an improved understanding of the $t\bar
t$-momentum distribution. Predictions for similar measurements at the
LHC are presented, demonstrating the sensitivity of the large rapidity
region both to the Standard Model contribution and effects from
''new physics''.
\end{quote}

\vspace*{\fill}
\begin{flushleft}
     December 12, 2011
\end{flushleft}
\end{titlepage}

\setcounter{footnote}{1}
\renewcommand{\thefootnote}{\fnsymbol{footnote}}


\section{Introduction}

Top quark production at hadron colliders is one of the most active
fields of current theoretical and experimental studies~\cite{Galtieri:2011yd}.
Theoretical predictions~\cite{Ahrens:2011px,Cacciari:2008zb,Moch:2008qy,Kidonakis:2010dk}
for the total production cross section
are in very good agreement with experimental results both at the Tevatron
at 1.96~TeV~\cite{Aaltonen:2010bs,Abazov:2009si} 
and the LHC at 7~TeV~\cite{ATLAStop,CMStop}. 
In contrast, sizable differences have been observed between theory 
predictions~\cite{Antunano:2007da,Kuhn:1998kw,Kuhn:1998jr,Bowen:2005ap} 
for the top quark charge asymmetry and measurements by the CDF and the 
D0 collaborations \cite{Abazov:2011rq,d043,d0,cdfcombined,cdfdilepton,Aaltonen:2011kc,cdf32,cdf19,Weinelt:2006mh,Hirschbuehl:2005bj,Schwarz:2006ud} 
at the Tevatron. 
The discrepancy is particularly pronounced for the subsample of
$t\bar t$ pairs with large invariant mass, $m_{t\bar t} > 450$~GeV, 
where a  $3.4 \sigma$ effect has been claimed~\cite{Aaltonen:2011kc}.
It is interesting to note, however, that the discrepancy is less prominent 
in the laboratory frame~\cite{Aaltonen:2011kc}. 
These discrepancies have triggered a large number of theoretical 
investigations, using these
results, either to restrict new physics like heavy 
axigluons~\cite{Ferrario:2009bz,Rodrigo:2010gm} or to
postulate a variety of new phenomena in 
the t-channel~\cite{Jung:2009jz,Cheung:2009ch,Shu:2009xf}
(see also ~\cite{Westhoff} for a recent review). At the same time the
robustness of the leading order QCD prediction has been studied 
in~\cite{Almeida:2008ug,Ahrens:2010zv},
where it has been argued that  next-to-leading (NLL) as well as
next-to-next-to leading (NNLL) logarithmic corrections do not
significantly modify the leading order result, in agreement with the
approach advocated in~\cite{Kuhn:1998kw,Kuhn:1998jr} 
(Note, however, the large corrections
observed in Ref.~\cite{Dittmaier:2008uj,Dittmaier:2007wz} 
for the corresponding studies of the
$t\bar t$+jet sample). The absence of large corrections in the asymmetry 
is at variance with the predictions based on Monte-Carlo simulations
where the numerator is evaluated in ${\cal O}(\alpha_s^3)$ and hence
leading order (LO), the denominator also in ${\cal O}(\alpha_s^3)$,
corresponding to terms of leading plus 
next-to-leading order. Inclusion of next-to-leading terms in the
denominator leads to a reduction by a factor roughly 0.7. 
A small modification of the
Standard Model (SM) prediction arises from inclusion of QED
corrections. In Ref.~\cite{Kuhn:1998kw} this effect was estimated 
to lead to an increase of the the asymmetry by a factor 1.09, in a recent 
analysis~\cite{Hollik:2011ps}, 
however, an enhancement factor of 1.2 has been obtained.
Obviously this small increase of the SM prediction for the asymmetry
cannot resolve the discrepancy between theory an experiment mentioned
above. 

In view of this ongoing discussion, a reanalysis of the SM prediction
seems appropriate. In this short note we evaluate the QED and weak
corrections to the asymmetry, confirming the results of 
Ref.~\cite{Hollik:2011ps}, and compare the SM result with the most recent 
measurements at Tevatron. 
Subsequently we study the effect of a cut on the transverse
momentum of the $t\bar t$ system on the asymmetry. 
A significant increase is observed, even for a cut as high as 20~GeV by
typically a factor $1.3$. This applies both at the parton and the 
hadronic level. Although the implications of this observation 
for the actual measurement can only be studied quantitatively by a 
(presently not available) full NNLO simulation, this cut
dependence, nevertheless, may serve as an indication of the sensitivity
of the asymmetry on details of the analysis. 

As noted already in~\cite{Kuhn:1998kw,Kuhn:1998jr}, 
a charge asymmetry may also be defined and
observed at the LHC. Since such an effect can only be observed in the
small subsample of $q\bar q$ induced events, specific kinematic regions
must be selected where gluon fusion is suppressed and $q\bar q$-
annihilation is enhanced. A particularly sensitive observable is the
ratio $A(Y)\equiv 
(N(y_t > y_{\bar t}) - N(y_t < y_{\bar t})/(N(y_t > y_{\bar t}) + N(y_t < y_{\bar t})$
with fixed average rapidity $Y\equiv (y_t+y_{\bar t})/2$. 
For large rapidities, say around $2$, the SM prediction
for $A_{t\bar t}(Y)$ amounts up to $0.05$ and might well be
detected at the LHC.

The quantity $A_{t\bar t}(Y)$ is also sensitive to physics beyond the SM. 
We use axigluons as one particularly illustrative example and study 
the sensitivity to amplitudes comparable in size to those suggested 
by recent Tevatron results.

\section{Amplitudes and partonic cross section}

\subsection{QCD asymmetry}

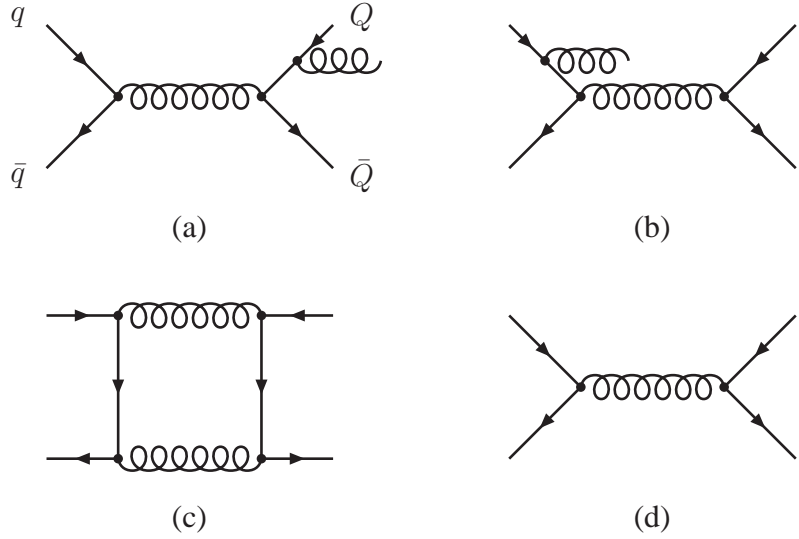
\begin{figure}[htb]
\begin{center}

\begin{picture}(200,200)(0,0)
\SetWidth{1.1}
\SetScale{.9}

\SetOffset(10,50)

\Vertex(30,30){2}
\Vertex(30,-30){2}
\Vertex(-30,30){2}
\Vertex(-30,-30){2}
\Gluon(-30,30)(30,30){5}{6}
\Gluon(30,-30)(-30,-30){5}{6}
\ArrowLine(60,30)(30,30)
\ArrowLine(30,30)(30,-30)
\ArrowLine(30,-30)(60,-30)
\ArrowLine(-60,30)(-30,30)
\ArrowLine(-30,30)(-30,-30)
\ArrowLine(-30,-30)(-60,-30)
\Text(0,-50)[]{(c)}

\SetOffset(185,50)

\Vertex(30,0){2}
\Vertex(-30,0){2}
\Gluon(-30,0)(30,0){5}{6}
\ArrowLine(-60,30)(-30,0)
\ArrowLine(-30,0)(-60,-30)
\ArrowLine(60,30)(30,0)
\ArrowLine(30,0)(60,-30)
\Text(0,-50)[]{(d)}

\SetOffset(10,160)

\Vertex(30,0){2}
\Vertex(-30,0){2}
\Vertex(45,15){2}
\Gluon(-30,0)(30,0){5}{6}
\Gluon(80,15)(45,15){5}{3}
\ArrowLine(-60,30)(-30,0)
\ArrowLine(-30,0)(-60,-30)
\ArrowLine(60,30)(45,15)
\Line(45,15)(30,0)
\ArrowLine(30,0)(60,-30)
\Text(-65,30)[]{$q$}
\Text(-65,-30)[]{$\bar q$}
\Text(65,30)[]{$Q$}
\Text(65,-30)[]{$\bar Q$}
\Text(0,-50)[]{(a)}

\SetOffset(185,160)

\Vertex(30,0){2}
\Vertex(-30,0){2}
\Vertex(-45,15){2}
\Gluon(-30,0)(30,0){5}{6}
\Gluon(-45,15)(-10,15){5}{3}
\ArrowLine(-60,30)(-45,15)
\Line(-45,15)(-30,0)
\ArrowLine(-30,0)(-60,-30)
\ArrowLine(60,30)(30,0)
\ArrowLine(30,0)(60,-30)
\Text(0,-50)[]{(b)}

\end{picture}
\end{center}
\caption{Origin of the QCD charge
asymmetry in hadroproduction of heavy quarks:
interference of final-state (a) with initial-state (b) gluon bremsstrahlung
plus interference of the box (c) with the Born diagram (d).
Crossed diagrams are omitted.
\label{fig:diagrams}}
\end{figure}

\begin{figure}[htb]
\begin{center}

\begin{picture}(200,100)(0,0)
\SetWidth{1.1}
\SetScale{.9}

\SetOffset(-10,50)

\Vertex(40,0){2}
\Vertex(70,0){2}
\Vertex(0,30){2}
\Vertex(0,-30){2}
\Vertex(-60,30){2}
\Vertex(-60,-30){2}
\Gluon(-60,30)(0,30){5}{6}
\Gluon(0,-30)(-60,-30){5}{6}
\Gluon(40,0)(70,0){4}{3}
\ArrowLine(0,30)(0,-30)
\ArrowLine(0,-30)(40,0)
\ArrowLine(40,0)(0,30)
\ArrowLine(100,30)(70,0)
\ArrowLine(70,0)(100,-30)
\ArrowLine(-90,30)(-60,30)
\ArrowLine(-60,30)(-60,-30)
\ArrowLine(-60,-30)(-90,-30)
\DashLine(25,30)(25,-30){5}
\DashLine(20,40)(-20,-40){5}
\Text(0,-60)[]{(a)}

\SetOffset(195,50)

\Vertex(40,0){2}
\Vertex(70,0){2}
\Vertex(0,30){2}
\Vertex(0,-30){2}
\Vertex(-60,30){2}
\Vertex(-60,-30){2}
\Gluon(-60,30)(0,-30){5}{8}
\Gluon(0,30)(-60,-30){5}{8}
\Gluon(40,0)(70,0){5}{3}
\ArrowLine(0,30)(0,-30)
\ArrowLine(0,-30)(40,0)
\ArrowLine(40,0)(0,30)
\ArrowLine(100,30)(70,0)
\ArrowLine(70,0)(100,-30)
\ArrowLine(-90,30)(-60,30)
\ArrowLine(-60,30)(-60,-30)
\ArrowLine(-60,-30)(-90,-30)
\DashLine(25,30)(25,-30){5}
\DashLine(20,40)(-20,-40){5}
\Text(0,-60)[]{(b)}

\end{picture}
\end{center}
\caption{Cut diagrams.\label{fig:cut}}
\end{figure}
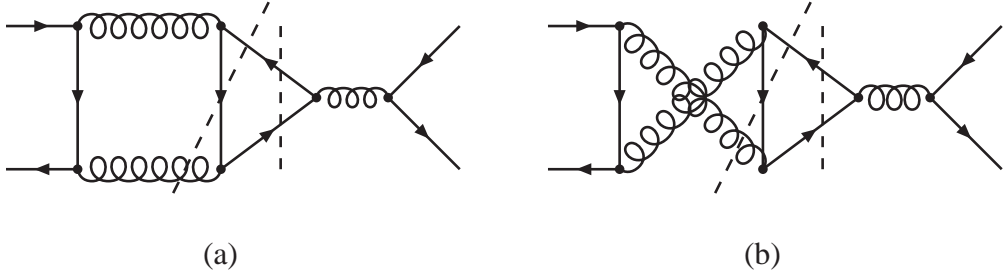

As shown in~\cite{Kuhn:1998kw,Kuhn:1998jr}, 
the dominant contribution to the charge asymmetry
originates from $q\bar{q}$ annihilation. Specifically, it originates 
from the interference between the Born amplitudes for 
$q\bar{q}\to Q\bar{Q}$ (Fig.~\ref{fig:diagrams}d) and the part of the one-loop 
correction, which is antisymmetric under the exchange of quark 
and antiquark (Fig.~\ref{fig:diagrams}c) (box and crossed box). 
To compensate the infrared divergences, this virtual correction 
must be combined with the interference between initial and 
final state radiation (Figs.~\ref{fig:diagrams}a,~\ref{fig:diagrams}b).
Diagrams with triple gluon coupling in both real and virtual 
corrections give rise to symmetric amplitudes~\cite{Kuhn:1998kw,Kuhn:1998jr}
and can be ignored. The corresponding contribution to the rate is conveniently 
expressed by the absorptive contributions (cuts) of the 
diagrams depicted in Fig~\ref{fig:cut}.

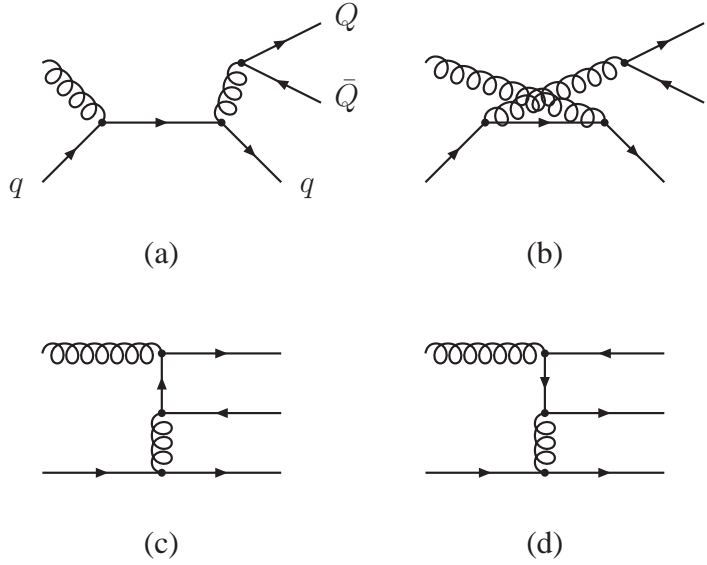
\begin{figure}[htb]
\begin{center}

\begin{picture}(280,200)(0,0)
\SetWidth{1.1}
\SetScale{.75}

\SetOffset(50,160)

\Vertex(30,0){2}
\Vertex(-30,0){2}
\Vertex(40,30){2}
\ArrowLine(-30,0)(30,0)
\Gluon(-60,30)(-30,0){5}{4}
\ArrowLine(-60,-30)(-30,0)
\Gluon(30,0)(40,30){5}{3}
\ArrowLine(30,0)(60,-30)
\ArrowLine(80,10)(40,30)
\ArrowLine(40,30)(80,50)
\Text(-55,-25)[]{$q$}
\Text(55,-25)[]{$q$}
\Text(70,40)[]{$Q$}
\Text(70,10)[]{$\bar Q$}
\Text(0,-50)[]{(a)}

\SetOffset(195,160)

\Vertex(30,0){2}
\Vertex(-30,0){2}
\Vertex(40,30){2}
\ArrowLine(-30,0)(30,0)
\Gluon(-60,30)(30,0){5}{10}
\ArrowLine(-60,-30)(-30,0)
\Gluon(-30,0)(40,30){5}{8}
\ArrowLine(30,0)(60,-30)
\ArrowLine(80,10)(40,30)
\ArrowLine(40,30)(80,50)
\Text(0,-50)[]{(b)}

\SetOffset(50,50)

\Vertex(0,30){2}
\Vertex(0,0){2}
\Vertex(0,-30){2}
\ArrowLine(60,0)(0,0)
\ArrowLine(0,0)(0,30)
\ArrowLine(0,30)(60,30)
\ArrowLine(-60,-30)(0,-30)
\ArrowLine(0,-30)(60,-30)
\Gluon(-60,30)(0,30){5}{7}
\Gluon(0,-30)(0,0){5}{3}
\Text(0,-50)[]{(c)}

\SetOffset(195,50)

\Vertex(0,30){2}
\Vertex(0,0){2}
\Vertex(0,-30){2}
\ArrowLine(60,30)(0,30)
\ArrowLine(0,30)(0,0)
\ArrowLine(0,0)(60,0)
\ArrowLine(-60,-30)(0,-30)
\ArrowLine(0,-30)(60,-30)
\Gluon(-60,30)(0,30){5}{7}
\Gluon(0,-30)(0,0){5}{3}
\Text(0,-50)[]{(d)}



\end{picture}
\end{center}
\caption{Origin of the QCD charge asymmetry in hadroproduction of heavy quarks
through flavor excitation.\label{fig:qg}}
\end{figure}

As a second contribution to the asymmetry we would like to mention 
``flavor excitation'' involving again antisymmetric interference 
terms of different amplitudes, in this case contributing to quark-gluon
scattering as shown in Fig.~\ref{fig:qg}
(amplitudes with triple gluon coupling again don't give rise 
to antisymmetric terms). Flavor excitation hardly contributes to 
the asymmetry at the Tevatron. At the LHC, however, it enhances 
the asymmetry in suitable chosen kinematical regions, as 
discussed in Ref.~\cite{Kuhn:1998jr}. 

Compact analytical results for the asymmetric parts of virtual 
plus soft radiation 
($E^g < E^g_{\rm cut}$) and of hard radiation ($E^g \ge E^g_{\rm cut}$) 
can be found in the Appendix of Ref.~\cite{Kuhn:1998kw}.

Let us recall that the color factors corresponding to Fig.~\ref{fig:cut}a 
and~\ref{fig:cut}b, after averaging over initial and summing over 
final states, are given by 
\bea
&& {\cal C}_a = \frac{1}{N_C^2} {\rm Tr}(T^a T^b T^c) \, {\rm Tr}(T^a T^c T^b)
= \frac{1}{16 \, N_C^2} (d_{abc}^2+f_{abc}^2)~, \nn \\
&& {\cal C}_b = \frac{1}{N_C^2} {\rm Tr}(T^a T^b T^c) \, {\rm Tr}(T^b T^c T^a) 
= \frac{1}{16 \, N_C^2} (d_{abc}^2-f_{abc}^2)~.
\eea
Without color factors the contributions to the differential cross-section 
from the two- and three-particle cuts in Fig.~\ref{fig:cut}a and
\ref{fig:cut}b are related by 
\beq
d\sigma_{a}(Q,\bar{Q})= -d\sigma_{b}(Q,\bar{Q})~.
\eeq
The asymmetric piece thus originates from the $d_{abc}^2$ term and, 
in leading order, its form is completely equivalent to the corresponding 
QED case. 


\subsection{QED asymmetry}

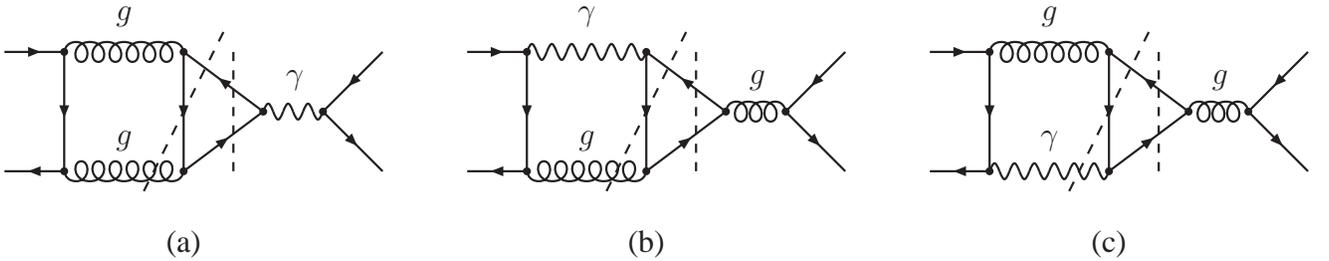
\begin{figure}[htb]
\begin{center}

\begin{picture}(380,100)(0,0)
\SetWidth{1.1}
\SetScale{.75}

\SetOffset(10,50)

\Vertex(40,0){2}
\Vertex(70,0){2}
\Vertex(0,30){2}
\Vertex(0,-30){2}
\Vertex(-60,30){2}
\Vertex(-60,-30){2}
\Gluon(-60,30)(0,30){5}{6}
\Gluon(0,-30)(-60,-30){5}{6}
\Photon(40,0)(70,0){4}{3}
\ArrowLine(0,30)(0,-30)
\ArrowLine(0,-30)(40,0)
\ArrowLine(40,0)(0,30)
\ArrowLine(100,30)(70,0)
\ArrowLine(70,0)(100,-30)
\ArrowLine(-90,30)(-60,30)
\ArrowLine(-60,30)(-60,-30)
\ArrowLine(-60,-30)(-90,-30)
\DashLine(25,30)(25,-30){5}
\DashLine(20,40)(-20,-40){5}
\Text(42,12)[]{$\gamma$}
\Text(-22.5,36)[]{$g$}
\Text(-22.5,-11)[]{$g$}
\Text(0,-50)[]{(a)}

\SetOffset(185,50)

\Vertex(40,0){2}
\Vertex(70,0){2}
\Vertex(0,30){2}
\Vertex(0,-30){2}
\Vertex(-60,30){2}
\Vertex(-60,-30){2}
\Photon(-60,30)(0,30){4}{6}
\Gluon(0,-30)(-60,-30){5}{6}
\Gluon(40,0)(70,0){5}{3}
\ArrowLine(0,30)(0,-30)
\ArrowLine(0,-30)(40,0)
\ArrowLine(40,0)(0,30)
\ArrowLine(100,30)(70,0)
\ArrowLine(70,0)(100,-30)
\ArrowLine(-90,30)(-60,30)
\ArrowLine(-60,30)(-60,-30)
\ArrowLine(-60,-30)(-90,-30)
\DashLine(25,30)(25,-30){5}
\DashLine(20,40)(-20,-40){5}
\Text(42,12)[]{$g$}
\Text(-22.5,36)[]{$\gamma$}
\Text(-22.5,-11)[]{$g$}
\Text(0,-50)[]{(b)}

\SetOffset(360,50)

\Vertex(40,0){2}
\Vertex(70,0){2}
\Vertex(0,30){2}
\Vertex(0,-30){2}
\Vertex(-60,30){2}
\Vertex(-60,-30){2}
\Gluon(-60,30)(0,30){5}{6}
\Photon(0,-30)(-60,-30){4}{6}
\Gluon(40,0)(70,0){5}{3}
\ArrowLine(0,30)(0,-30)
\ArrowLine(0,-30)(40,0)
\ArrowLine(40,0)(0,30)
\ArrowLine(100,30)(70,0)
\ArrowLine(70,0)(100,-30)
\ArrowLine(-90,30)(-60,30)
\ArrowLine(-60,30)(-60,-30)
\ArrowLine(-60,-30)(-90,-30)
\DashLine(25,30)(25,-30){5}
\DashLine(20,40)(-20,-40){5}
\Text(42,12)[]{$g$}
\Text(-22.5,36)[]{$g$}
\Text(-22.5,-11)[]{$\gamma$}
\Text(0,-50)[]{(c)}

\end{picture}
\end{center}
\caption{Representative diagrams contributing to the QCD-QED
interference term.\label{fig:gamma}}
\end{figure}

Already at this point we would like to discuss the closely 
related QED contribution to the asymmetry. Let us 
start with diagram shown in Fig.~\ref{fig:gamma}a
following again Ref.~\cite{Kuhn:1998kw}. 
The QCD box leads to a color octet and color singlet configuration, 
and the latter interferes with $t\bar{t}$ production through the photon. 
A similar consideration applies to interference between initial 
and final state radiation. These two contributions are indicated 
in Fig.~\ref{fig:gamma}a by the two cuts, and in combination 
lead to an additional asymmetric term which can be obtained 
from the QCD asymmetry through the 
replacement~\cite{Kuhn:1998kw,Kuhn:1998jr}
\beq
\frac{\alpha_S}{2} \, \left( \frac{d_{abc}^2}{4}\right)^2 \to 
\alpha_{\rm QED} \, Q_t \, Q_q~.
\label{eq:QCDQED}
\eeq
Another QED term originates from the interference between the gluon-$\gamma$
box with the QCD Born amplitude. Since gluons and photon are distinct 
fields, two contributions as depicted in Fig.~\ref{fig:gamma}b
and~\ref{fig:gamma}c arise~\footnote{These small terms had been 
neglected in~\cite{Kuhn:1998jr}, in~\cite{Kuhn:1998kw} only 
one of the two had been included. The present result is in 
agreement with~\cite{Hollik:2011ps}}. Each of these contributes 
with the factor given in \Eq{eq:QCDQED}.
In total the relative factor between QCD and QED asymmetries amounts to 
\beq
f_q^{\rm QED} = 3 \, \frac{\alpha_{\rm QED} \, Q_t \, Q_q}
{\displaystyle \frac{\alpha_S}{2} \, \left( \frac{d_{abc}^2}{4}\right)^2}
= \frac{\alpha_{\rm QED}}{\alpha_S} \, \frac{36}{5} \, Q_t \, Q_q
\label{eq:fqQED}
\eeq
for one quark species. Let us, in a first step, assume identical 
functional dependence of the PDFs for $u$ and $d$ valence quarks 
in the proton (modulo the obvious factor two) and similarly for antiquarks 
in the antiproton. Assuming, furthermore, dominance of valence 
quark-antiquark annihilation in $t\bar t$ production, 
the relative contributions of the $u\bar u$ versus $d\bar d$ induced 
reactions to the cross section have to be weighted with the ratio 4:1. 
The QED asymmetry has to be weighted, furthermore, with relative 
factors $f_u^{\rm QED}$ and $f_d^{\rm QED}$ respectively. 
The relative QED contribution thus amounts to
\beq
f^{\rm QED}_{\rm Tevatron} = \frac{4 f_u^{\rm QED} +  f_d^{\rm QED}}{5} = 
\frac{\alpha_{\rm QED}}{\alpha_S} \, \frac{56}{25} \approx 0.18~,
\label{eq:QED}
\eeq
at the Tevatron, and thus to an enhancement of nearly twenty percent 
of the QCD asymmetry, in good agreement with the more detailed 
numerical studies presented below and with the results 
of~\cite{Hollik:2011ps}.
Compared to proton-antiproton collisions the relative importance 
of $u\bar u$ versus $d\bar d$ annihilation at the LHC is shifted from 
approximately $4:1$ to $2:1$, thus reducing $f^{\rm QED}$ to
$f^{\rm QED}_{\rm LHC} = (2 f^{\rm QED}_u + f^{\rm QED}_d)/3 \approx 0.13$, 
which is lower than the result of~\Eq{eq:QED} by a factor $5/7$. 
The results using standard PDFs are close to these values and will be listed 
in Sect.~\ref{sec:tevatron}.

\subsection{Weak asymmetry}

\begin{figure}[htb]
\begin{center}

\begin{picture}(380,100)(0,0)
\SetWidth{1.1}
\SetScale{.75}

\SetOffset(10,50)

\Vertex(40,0){2}
\Vertex(70,0){2}
\Vertex(0,30){2}
\Vertex(0,-30){2}
\Vertex(-60,30){2}
\Vertex(-60,-30){2}
\Gluon(-60,30)(0,30){5}{6}
\Gluon(0,-30)(-60,-30){5}{6}
\Photon(40,0)(70,0){4}{3}
\ArrowLine(0,30)(0,-30)
\ArrowLine(0,-30)(40,0)
\ArrowLine(40,0)(0,30)
\ArrowLine(100,30)(70,0)
\ArrowLine(70,0)(100,-30)
\ArrowLine(-90,30)(-60,30)
\ArrowLine(-60,30)(-60,-30)
\ArrowLine(-60,-30)(-90,-30)
\DashLine(25,30)(25,-30){5}
\DashLine(20,40)(-20,-40){5}
\Text(42,12)[]{$Z$}
\Text(-22.5,36)[]{$g$}
\Text(-22.5,-11)[]{$g$}
\Text(0,-50)[]{(a)}

\SetOffset(185,50)

\Vertex(40,0){2}
\Vertex(70,0){2}
\Vertex(0,30){2}
\Vertex(0,-30){2}
\Vertex(-60,30){2}
\Vertex(-60,-30){2}
\Photon(-60,30)(0,30){4}{6}
\Gluon(0,-30)(-60,-30){5}{6}
\Gluon(40,0)(70,0){5}{3}
\ArrowLine(0,30)(0,-30)
\ArrowLine(0,-30)(40,0)
\ArrowLine(40,0)(0,30)
\ArrowLine(100,30)(70,0)
\ArrowLine(70,0)(100,-30)
\ArrowLine(-90,30)(-60,30)
\ArrowLine(-60,30)(-60,-30)
\ArrowLine(-60,-30)(-90,-30)
\DashLine(25,30)(25,-30){5}
\DashLine(20,40)(-20,-40){5}
\Text(42,12)[]{$g$}
\Text(-22.5,36)[]{$Z$}
\Text(-22.5,-11)[]{$g$}
\Text(0,-50)[]{(b)}

\SetOffset(360,50)

\Vertex(40,0){2}
\Vertex(70,0){2}
\Vertex(0,30){2}
\Vertex(0,-30){2}
\Vertex(-60,30){2}
\Vertex(-60,-30){2}
\Gluon(-60,30)(0,30){5}{6}
\Photon(0,-30)(-60,-30){4}{6}
\Gluon(40,0)(70,0){5}{3}
\ArrowLine(0,30)(0,-30)
\ArrowLine(0,-30)(40,0)
\ArrowLine(40,0)(0,30)
\ArrowLine(100,30)(70,0)
\ArrowLine(70,0)(100,-30)
\ArrowLine(-90,30)(-60,30)
\ArrowLine(-60,30)(-60,-30)
\ArrowLine(-60,-30)(-90,-30)
\DashLine(25,30)(25,-30){5}
\DashLine(20,40)(-20,-40){5}
\Text(42,12)[]{$g$}
\Text(-22.5,36)[]{$g$}
\Text(-22.5,-11)[]{$Z$}
\Text(0,-50)[]{(c)}

\end{picture}
\end{center}
\caption{Representative diagrams contributing to the QCD-weak
interference term.\label{fig:Z}}
\end{figure}
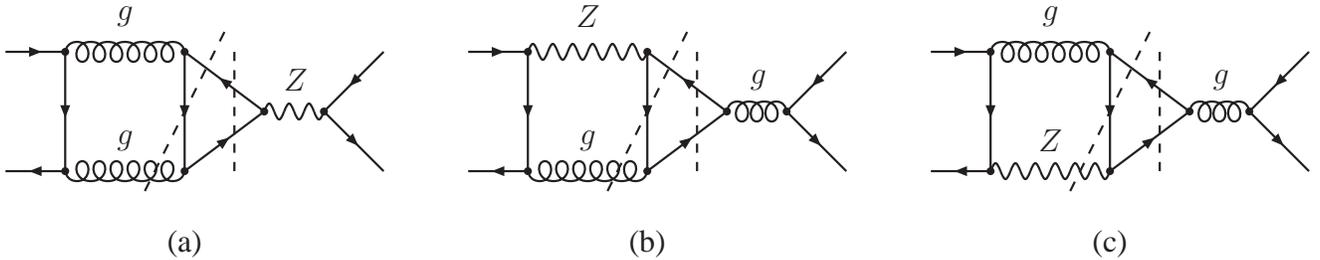

Weak and electromagnetic interactions are of comparable strength at 
energies characteristic for the Tevatron and the LHC. 
Hence, contributions similar to those depicted in 
Figs.~\ref{fig:gamma}a,~\ref{fig:gamma}b and~\ref{fig:gamma}c 
with the photon replaced by the 
$Z$ boson should be considered at the same footing. Let us 
start with the contribution depicted in Fig.~\ref{fig:Z}a,
where the $Z$ boson is off-shell with virtuality $\hat s \gg m_Z^2$.
The result is obtained~\cite{Kuhn:1998jr} from the photon  
contribution through the replacement 
\beq
Q_t \, Q_q \to \frac{(2\, I_t-4\,  Q_t\, s_W^2)(2\, I_q-4\,  Q_q\, s_W^2)}
{16\, s_W^2 \, c_W^2} \, \frac{1}{1-m_Z^2/\hat s}~,
\eeq
with $s_W^2$ and $c_W^2$ denoting the squares of the sine and cosine of the weak 
mixing angle, respectively, and $I_q$ the weak isospin of the relevant quark. 
Adopting the weighted average similar to~\Eq{eq:QED} we find
\beq
f_1^{\rm weak} = \frac{\alpha_{\rm QED}}{\alpha_S}\, \frac{36}{5} \,
\frac{1-\frac{8}{3}\, s_W^2}{1-m_Z^2/\hat s} \, \frac{1}{16\, s_W^2 \, c_W^2} \,
\frac{4(1-\frac{8}{3}\, s_W^2)+(-1+\frac{4}{3}\, s_W^2)}{5} \approx
4.4 \times 10^{-3}~,
\eeq
for the contribution Fig.~\ref{fig:Z}a. Note that, as a consequence
of the cancellation between up and down quarks, and the smallness of the 
weak coupling, this result is smaller by more than a factor $10$ than 
the corresponding photonic result. For proton-proton collision $f_1^{\rm weak}$
is further reduced down to $7 \times 10^{-4}$ and thus is completely 
negligible.

It is tempting to estimate the contributions from diagrams~\ref{fig:Z}b
and \ref{fig:Z}c along the same lines. Independently of any detailed 
considerations the same compensation between $u$- and $d$-quark 
contributions will arise, and in the limit $m_Z^2 \ll \hat s$, 
and assuming that final states with real $Z$ radiation are included 
in the sample of $t\bar t$ events, the analogs of Figs.~\ref{fig:Z}b 
and~\ref{fig:Z}c will contribute identically to 
Fig.~\ref{fig:Z}a, enhancing the correction from $0.5\%$ to 
$1.5\%$. Alternatively, one may perform an explicit calculation of 
Figs.~\ref{fig:Z}b and~\ref{fig:Z}c, allowing for a 
separation of real and virtual $Z$ boson radiation.

Let us, finally mention that contributions to the asymmetry involving 
the squared electroweak amplitude $q\bar q \stackrel{\gamma, Z}{\to} t\bar t$
are of order $\alpha_{\rm QED}^2/\alpha_S^2$ and thus at most 
of ${\cal O}(1\%)$. 
Furthermore, if we would include terms of order $\alpha_{\rm QED}^2/\alpha_S^2$
into consideration, terms of order $\alpha_S^2 \, \alpha_{\rm QED}$
in the total cross-section, i.e. electroweak corrections to $t\bar t$
production (leading to corrections of comparable 
size~\cite{Kuhn:2006vh,Kuhn:2005it,Bernreuther:2005ej,Bernreuther:2006vg,Hollik:2007sw}
should be included as well. 
Since even the NLO QCD corrections to the asymmetry 
of ${\cal O}(\alpha_S)$ have not been evaluated to date, 
it seems unnecessary to include these ${\cal O}(\alpha_{\rm QED}^2)$
terms into consideration.

Indeed, considering the large positive NLO QCD corrections to the 
cross-section amounting to about $30\%$ at the Tevatron, it 
seems plausible to assign a comparable uncertainty to the asymmetry. 
However, in agreement 
with~\cite{Kuhn:1998kw,Kuhn:1998jr,Almeida:2008ug,Ahrens:2010zv}, 
we shall assume that 
the central value of the asymmetry will not be shifted by the large 
corrections, in other words, that symmetric and antisymmetric 
parts of the cross-section are shifted by the same factor.


\begin{figure}[ht]
\begin{center}
\includegraphics[width=8cm]{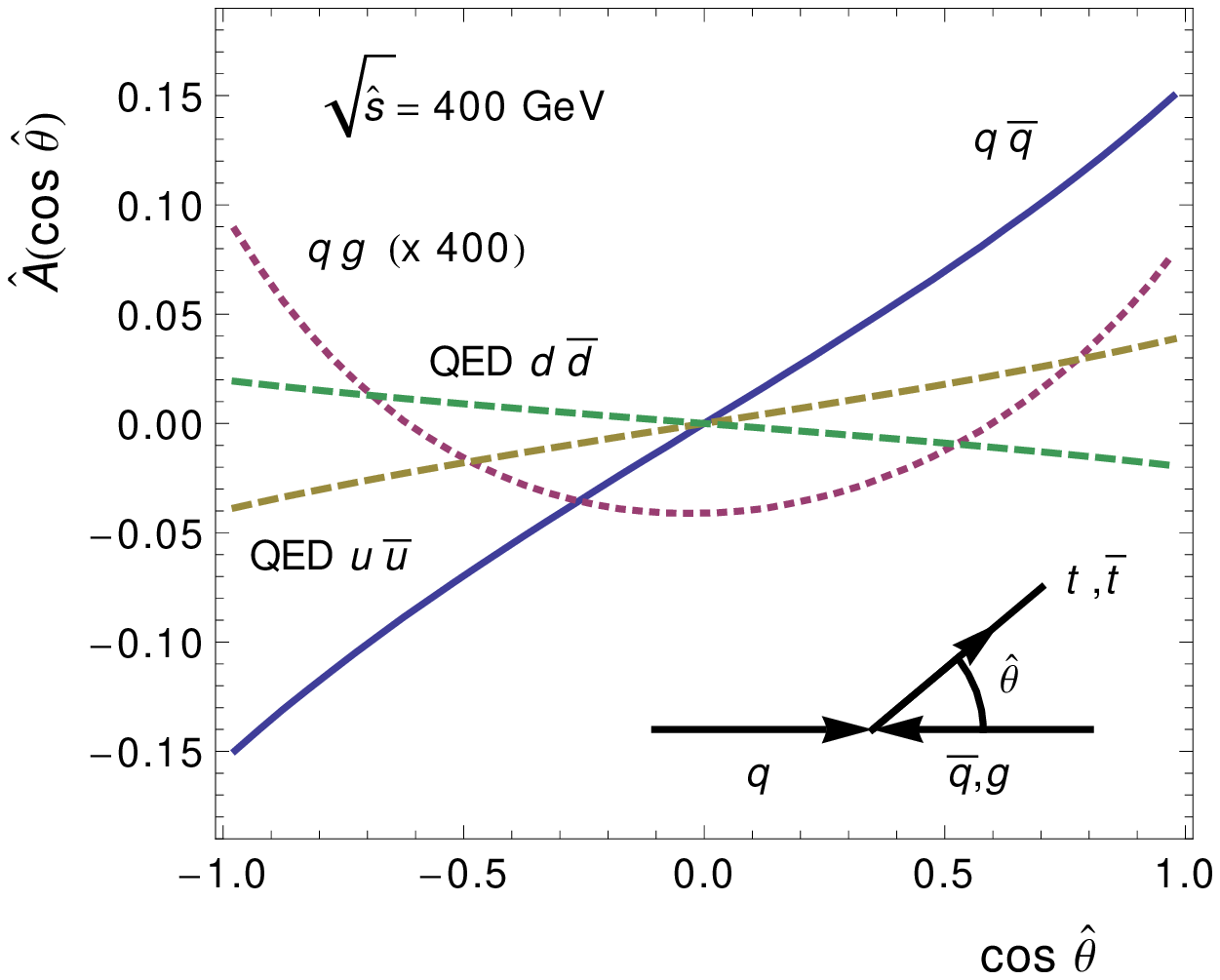}
\includegraphics[width=8cm]{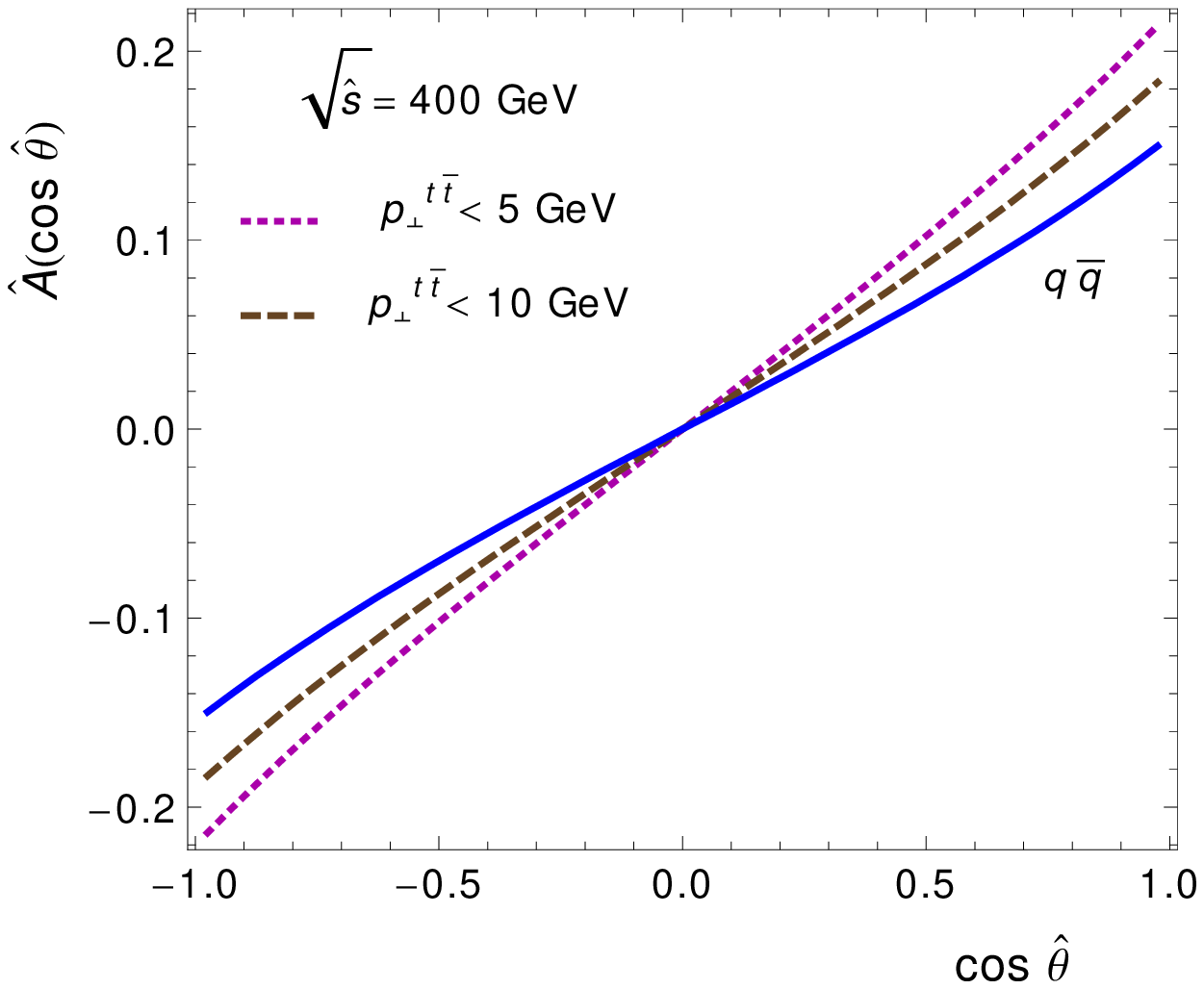}
\includegraphics[width=8cm]{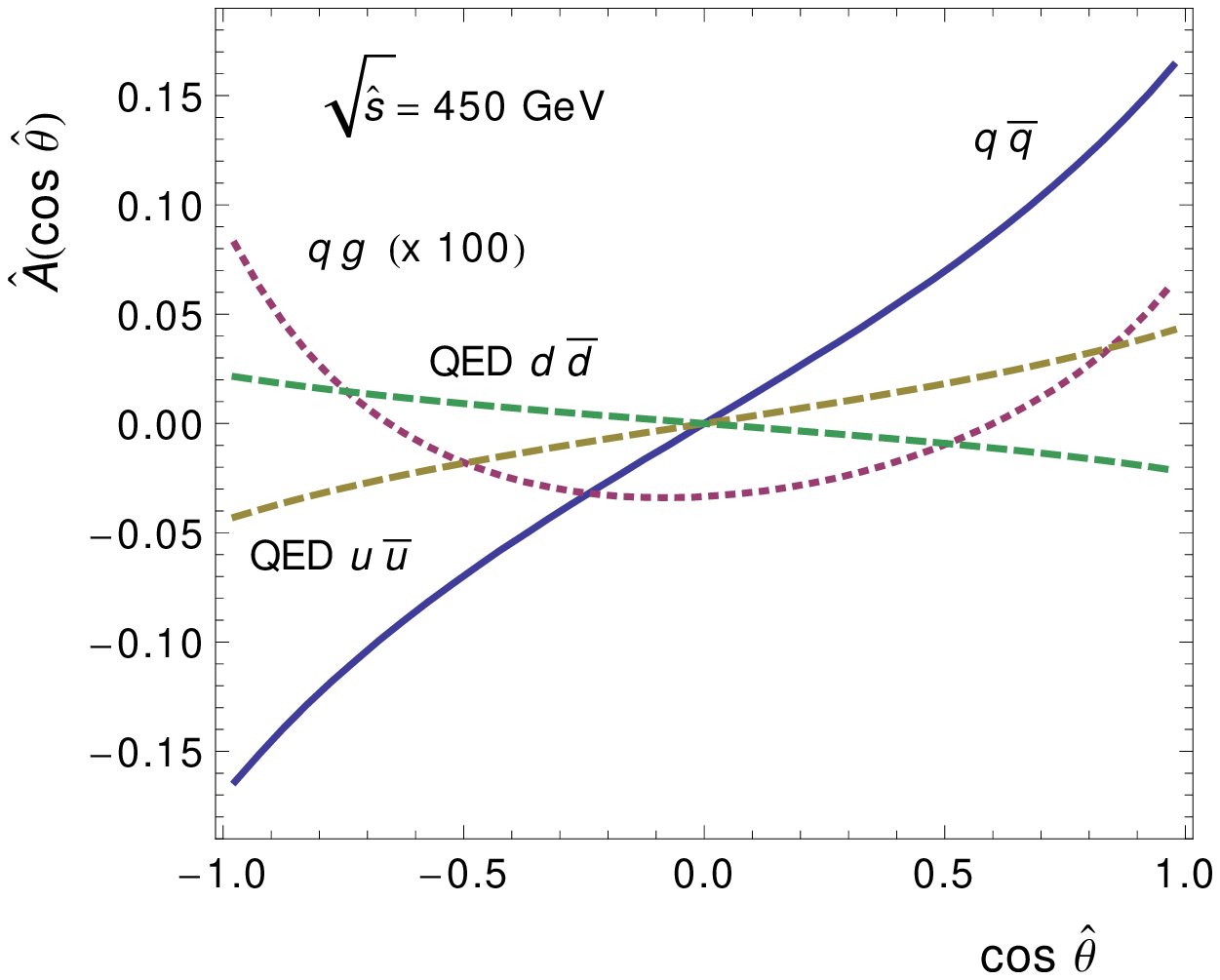}
\includegraphics[width=8cm]{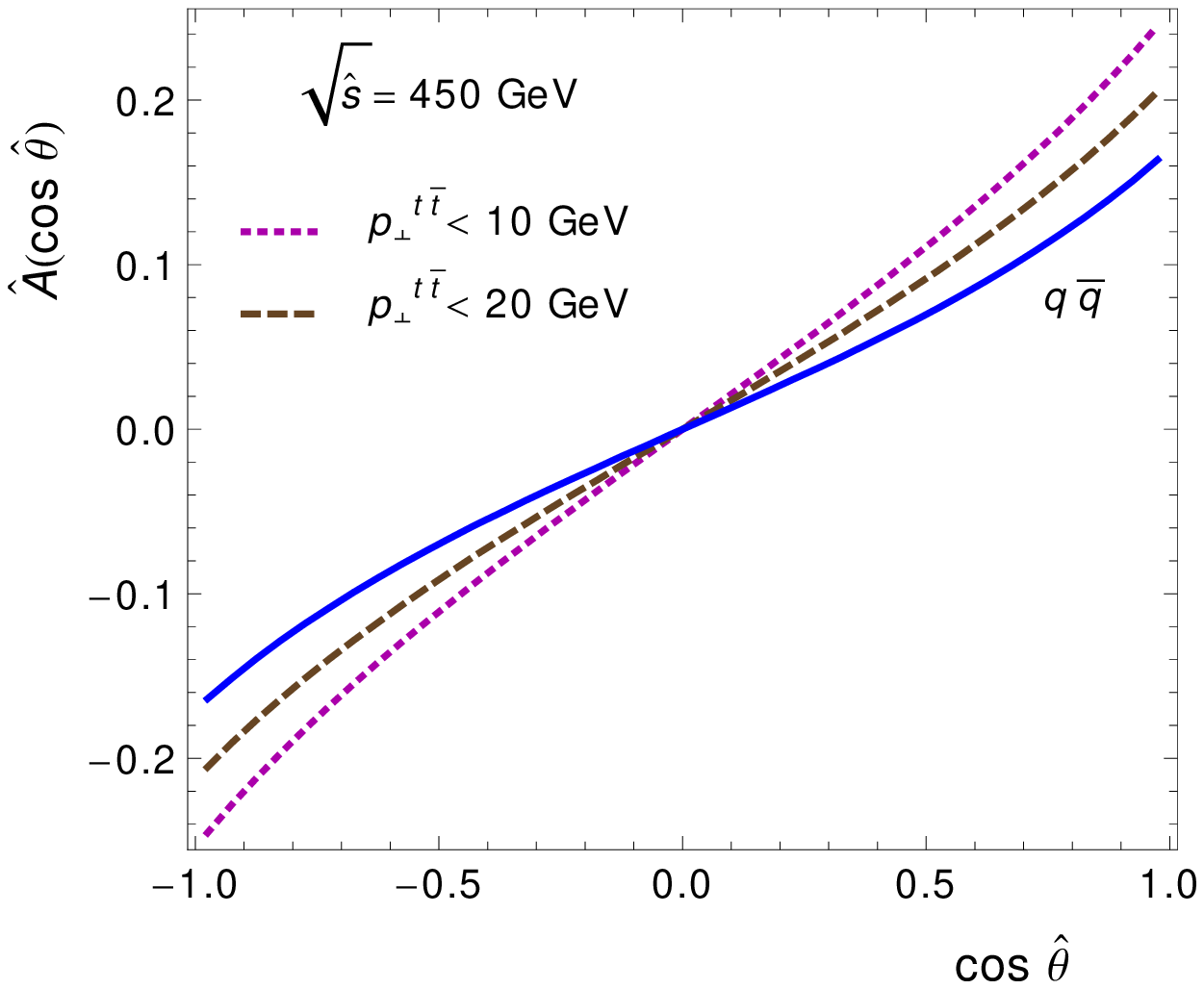}
\includegraphics[width=8cm]{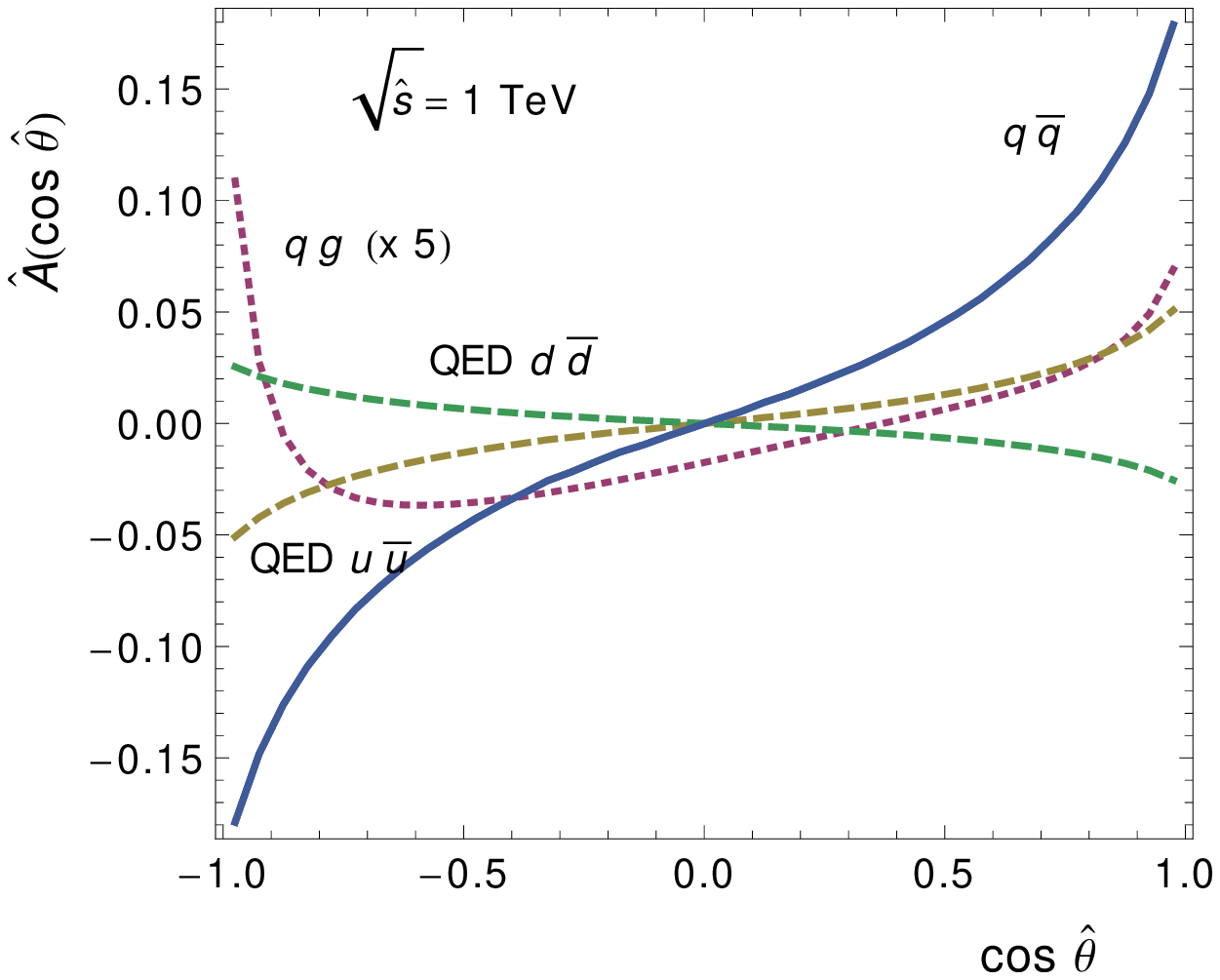}
\includegraphics[width=8cm]{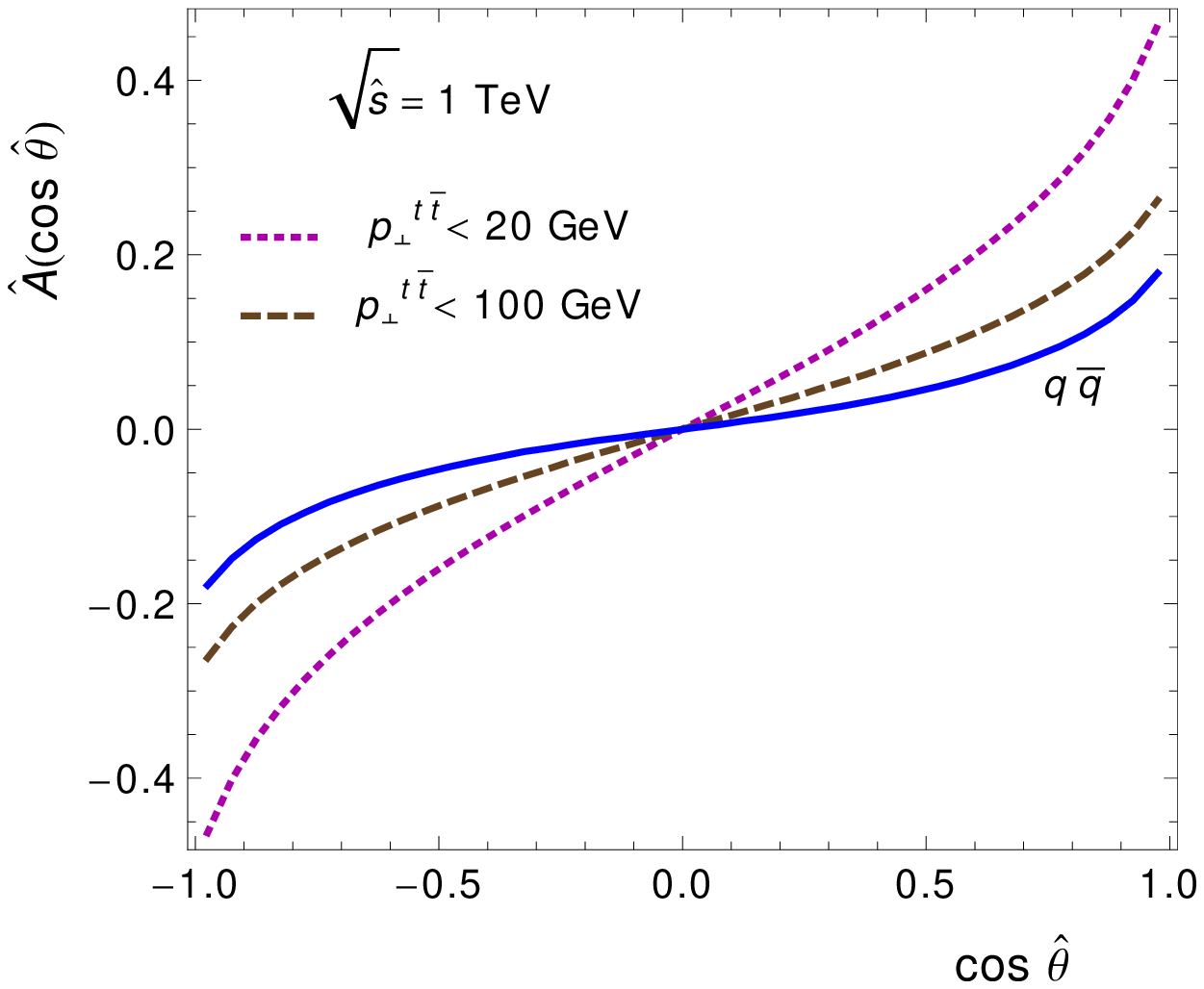}
\caption{Left plots: differential partonic asymmetries from $q\bar q$ 
and $qg$ induced events for different 
choices of the partonic center of mass energy. 
The asymmetry from mixed QED-QCD contributions
is shown separately for $u$ and $d$ quarks. 
Right plots: differential partonic asymmetries after 
introducing a cut in the transverse momentum of the 
$t\bar t$ pair $p_\perp^{t\bar t}$. 
\label{fig:partonic}}
\end{center}
\end{figure}

\subsection{Numerical results at the partonic level}

Results for the angular distribution of the antisymmetric part of the 
cross section are shown in Fig.~\ref{fig:partonic} (left plots) 
for fixed partonic center of mass energies, and divided by the 
total $q\bar q$ induced cross-section:
\beq
\hat A (\cos \hat \theta) = \frac{1}{\sigma_{qq}} \, 
\frac{d\sigma_A(\cos \hat \theta)}{d \cos \hat \theta}~.
\eeq
QED induced terms are shown
separately for $u$ and $d$ quarks, as well as the partonic 
asymmetry generated by $qg$ collisions. 
The QED generated asymmetries for $u$ and $d$ quarks follows the 
proportionality factor in \Eq{eq:fqQED}, amounting to 
$\alpha_{\rm QED}/\alpha_S \, \frac{36}{5} \, Q_t \, Q_q \approx 0.38 \, Q_q$. 
The piece generated by $qg$ collisions (flavor excitation) remains
small in the whole kinematic region of interest.

At this point it is instructive to study the effect of cuts
on gluon emission, which lead to an enhancement of the asymmetry. 
To understand the origin of this phenomenon, let us in a first 
step recall that the inclusive asymmetry is positive, which 
can be qualitatively understood from the requirement that 
the final state configuration with minimal modification of the color field 
is favored, corresponding to minimal change of the direction
of the color source. On the other hand, if one request a hard gluon 
in the final state, this configuration is more probable for backward going 
top quarks, thus emitting hard bremsstrahlung. 
Hence, for events with tagged hard gluons a negative 
charge asymmetry is expected.
Conversely, one expects a sizable increase of the inclusive asymmetry, 
if one cuts on events with real gluon emission. 
This is demonstrated in Fig.~\ref{fig:partonic} (right plots) where the 
dashed and the dotted curves are obtained for a cut on the transverse 
momentum of the top quark pair $p_\perp^{t\bar t} < p_\perp^{\rm max}$
with $p_\perp^{\rm max}$ chosen differently for different $\hat s$.
We observe an increase of the asymmetry by a significant amount. 
The effect is strongly dependent on the precise value of $p_\perp^{\rm max}$
and becomes more pronounced at larger $\hat s$, a consequence of the 
relatively larger amount of real radiation in the fully inclusive 
sample.

\begin{figure}[ht]
\begin{center}
\includegraphics[width=8cm]{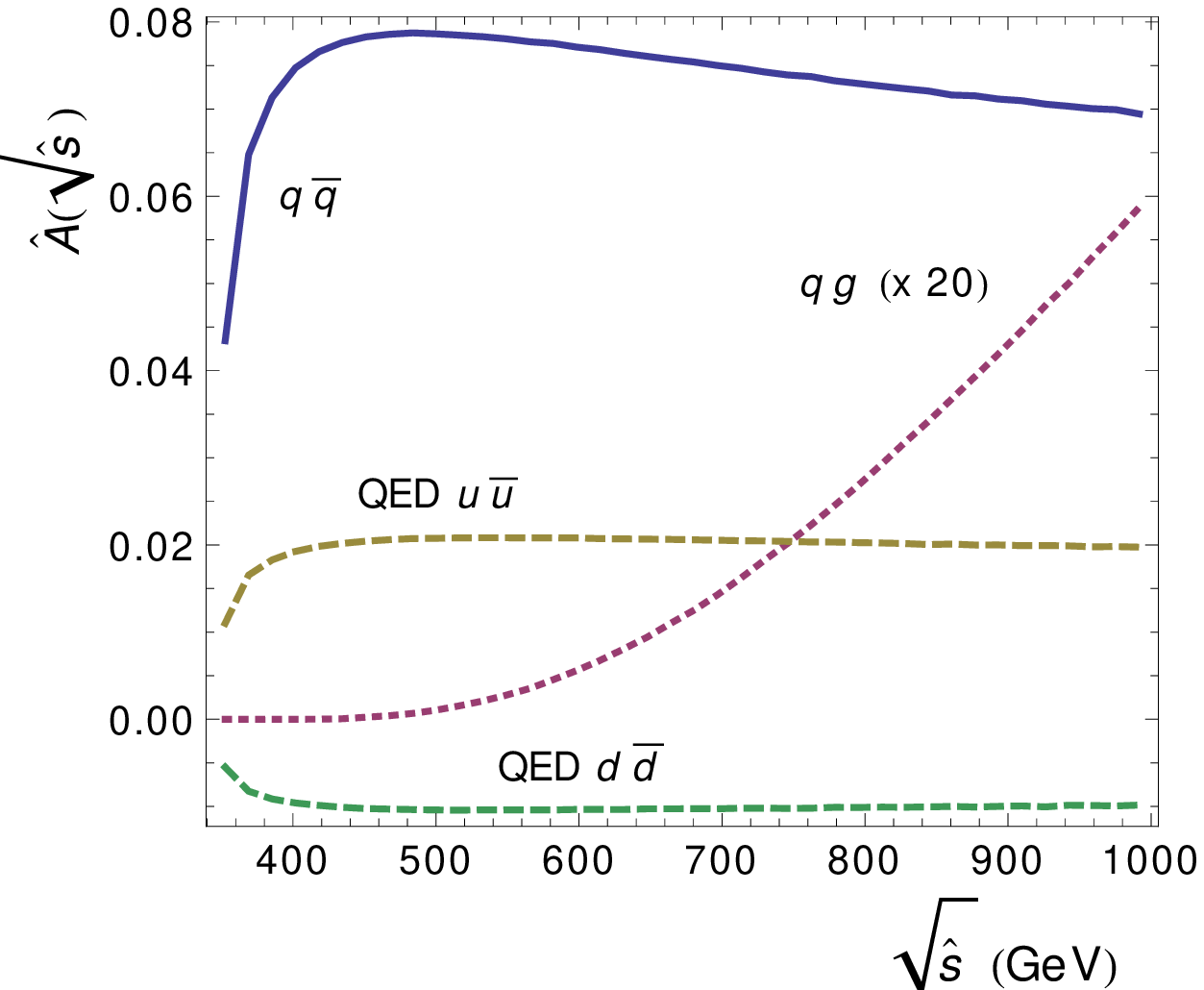}
\includegraphics[width=8cm]{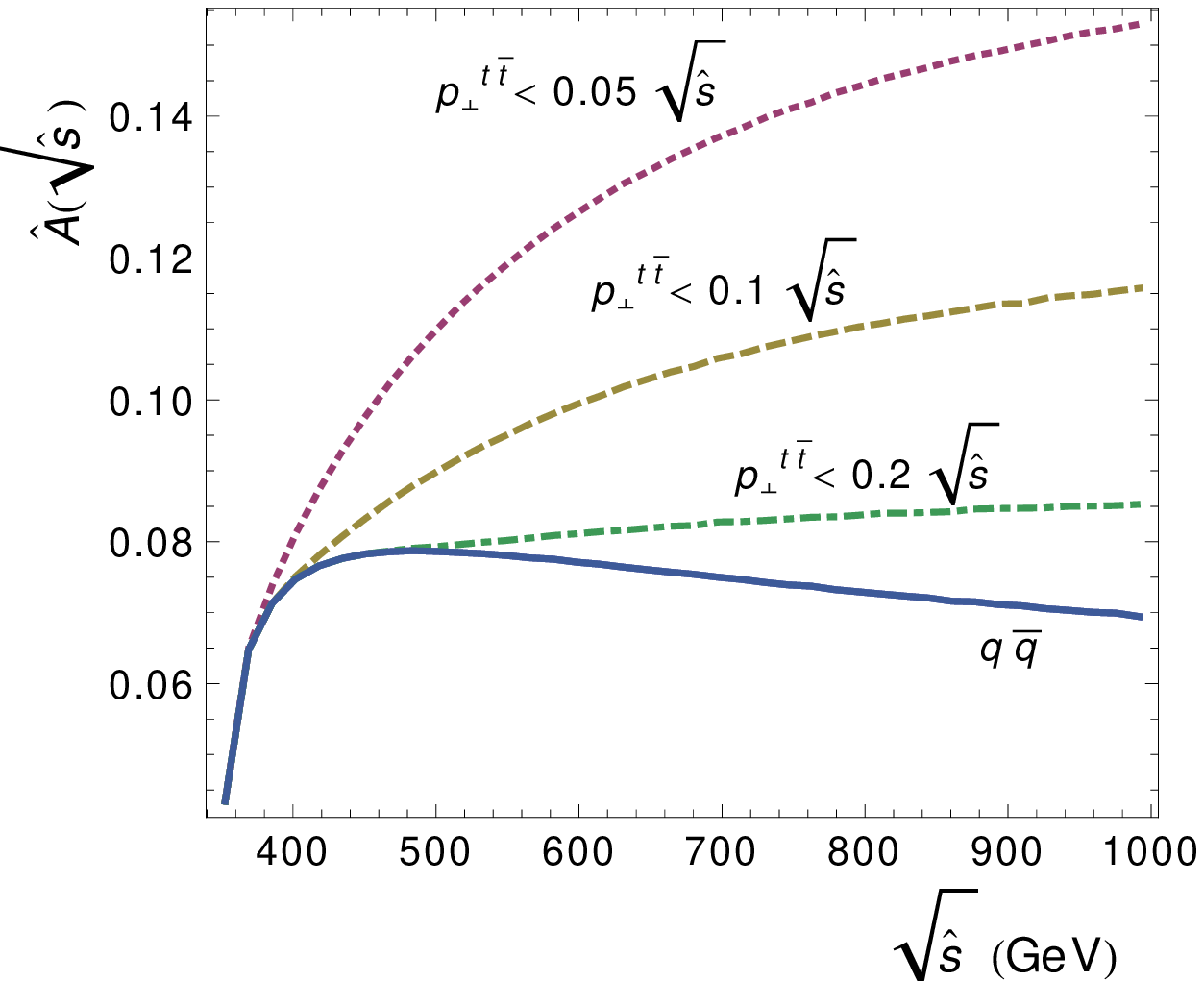}
\caption{Left plot: integrated partonic asymmetry from $q\bar q$ 
and $qg$ induced events for different 
choices of the partonic center of mass energy. 
The asymmetry from mixed QED-QCD contributions
is shown separately for $u$ and $d$ quarks. 
Right plot: integrated partonic asymmetry after 
introducing a cut in the transverse momentum of 
the $t\bar t$ pair $p_\perp^{t\bar t}$. 
\label{fig:partonicshat}}
\end{center}
\end{figure}

The energy dependence of the integrated asymmetry
(again normalized relative to the Born cross-section
$\sigma(q\bar q \to t\bar t$), 
\beq
\hat A = \int_0^1 \hat A (\cos \hat \theta) \, d\cos \hat \theta
-\int_{-1}^0 \hat A (\cos \hat \theta) \, d\cos \hat \theta
\eeq
is displayed in Fig.~\ref{fig:partonicshat}. 
As in Fig.~\ref{fig:partonic}, the left plot in Fig.~\ref{fig:partonicshat} 
show the pure QCD asymmetry generated by $q\bar q$ and $qg$ 
events, and the QCD-QED mixed asymmetry for $u$ and $d$ initial quarks. 
A rapid increase of the asymmetry is observed in the region 
very close to threshold, a consequence of the S-wave--P-wave interference 
of the asymmetry. 
The right plot in Fig.~\ref{fig:partonicshat} illustrates the effect 
on the QCD induced asymmetry from introducing a cut 
on the transverse momentum of the top quark 
pair $p_\perp^{t\bar t}$. Again a sizable increase of $\hat A$ is observed, 
which depends strongly on the choice of the cut.  

We use the following values for the top quark mass
$M_t = 173.3 (1.1)~{\rm GeV}$~\cite{:1900yx}, the strong coupling
$\alpha_S(M_Z) = 0.1184 (7)$~\cite{Nakamura:2010zzi}, 
the QED coupling $\alpha_{\rm QED}(M_Z) = 1/127$, and 
the square of the sine of the weak mixing $s_W^2=0.23$.
The renormalization scale in Fig.~\ref{fig:partonic} and
Fig.~\ref{fig:partonicshat} has been set to the partonic center 
of mass energy, $\mu=\sqrt{\hat{s}}$.

\section{Hadronic collisions}

\subsection{Generalities}

The asymmetry can, in principle, be studied in more detail by measuring 
in addition $m_{t\bar t}$, the invariant mass of the $t\bar t$ system, 
and its transverse momentum, which is balanced by the transverse momentum 
of an additionally radiated gluon. 
Controlling both observables together with the asymmetry 
would provide detailed information on the production 
dynamics. Although such a measurement is only possible with large
statistics, a first step into this direction has been performed 
by the CDF collaboration by separating events with invariant 
mass of the $t\bar t$-system above and below $450$~GeV. However, 
for a complete investigation the statistical precision is far 
too small. 

In the following we will present in a first step, our 
predictions for the Tevatron, subsequently for the LHC.
We will include the update on the electromagnetic and weak 
corrections, discuss the implications of a $p_\perp^{t \bar t}$
cut on the asymmetry, and compare our results to the most 
recent experimental results.
In a second step we present detailed predictions for the LHC
and identify kinematic regions where the charge asymmetry 
could be observed. In view of the indications at the Tevatron for 
a sizable excess of the asymmetry, we then investigate the 
implications for the charge asymmetry at the LHC in two 
benchmark scenarios beyond the SM. 

Different choices of parton distribution 
functions of the MSTW2008 set~\cite{Martin:2009iq} are used
to obtain the theoretical predictions of the asymmetry in 
hadronic collisions, specifically we consider MSTW2008LO and MSTW2008NLO. 
The dependence of the asymmetry on the choice of PDFs is, however, small.
The factorization and renormalization scales are varied between 
$\mu=m_t/2$ and $\mu=\sqrt{\hat s}$. 
This dependence gives the bulk of the estimated theoretical 
error because the asymmetry if proportional to $\alpha_S(\mu)$. 
The variation of the top mass within its experimental error 
is also considered; its effect on the asymmetry is also small.

\subsection{Tevatron}

\label{sec:tevatron}

\begin{figure}[t]
\begin{center}
\includegraphics[width=8cm]{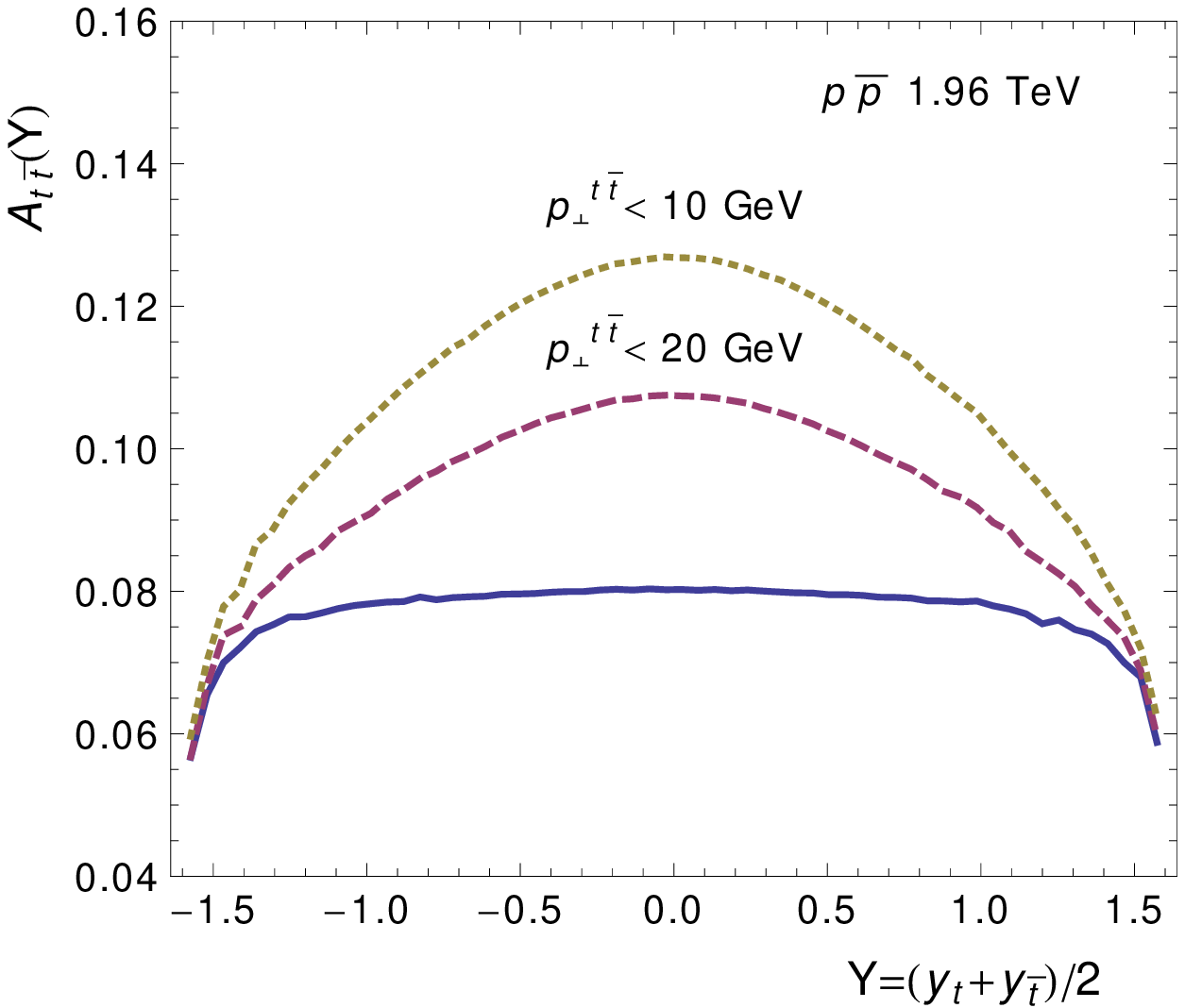}
\caption{Rapidity dependence of the asymmetry $A_{t\bar t} (Y)$
as defined in~\Eq{eq:pair}. Solid line: no cut on $p_\perp^{t\bar t}$, 
dotted/dashed line  $p_\perp^{\rm max} = 10$~GeV / $20$~GeV.
\label{fig:AY}}
\end{center}
\end{figure}

Let us start with our updated predictions for the Tevatron. 
Assuming that the rapidities of $t$ and $\bar t$ have 
been measured simultaneously, one defines the asymmetry
\beq
A_{t\bar t}\, (Y)=\frac{N(y_t>y_{\bar t})-N(y_{\bar t}>y_t)}
{N(y_t>y_{\bar t})+N(y_{\bar t}>y_t)}~,
\label{eq:pair}
\eeq
where $Y=(y_t+y_{\bar t})/2$ has been fixed. 
The results as a function of $Y$ are shown in Fig.~\ref{fig:AY}. 
An almost flat $A_{t\bar t} (Y)$ of around $8\%$ is observed. 
Two versions of the integrated asymmetry have been introduced
in Refs.~\cite{Antunano:2007da,Kuhn:1998kw,Kuhn:1998jr}: 
the forward--backward asymmetry in the laboratory frame
\beq
A_{\rm lab}=\frac{N(y_t>0)-N(y_t<0)}{N(y_t>0)+N(y_t<0)}
= \frac{N(y_t>0)-N(y_{\bar t}>0)}{N(y_t>0)+N(y_{\bar t}>0)}~,
\eeq
and the asymmetry in the $t\bar t$ rest frame
\beq
A_{t\bar t}=\frac{N(y_t>y_{\bar t})-N(y_{\bar t}>y_t)}
{N(y_t>y_{\bar t})+N(y_{\bar t}>y_t)}~.
\eeq
Results for both of them are listed in Table~\ref{tab:Attbar}. 
This Table also lists separately the contributions from pure 
QCD and from QED, the latter again separated for up and down quark. 
The relative QED contribution (QED~$u\bar u$~+~QED~$d\bar d$)~/~QCD 
as obtained from Table~\ref{tab:Attbar} is close to 0.18 for all the 
cases considered and thus in agreement with the expectations based 
on~\Eq{eq:QED}.
Similarly, we also give the results for the weak terms in 
the asymmetry, using the approximation $m_Z^2 \ll \hat s$. 
Note that the weak contribution is strongly suppressed, even 
in comparison with the QED piece, and hence can be safely 
treated in this approximation. 
The relative contribution of $gg$, $u\bar u$ and $d\bar d$
initiated top quarks is also listed in this Table. 
Due to the deviation of the relative amount of  
$u$- and $d$- contributions from the simple approximation 
$4:1$ we find a slight deviation from~\Eq{eq:QED}.
The overall factor $1.21$ is consistent with~\cite{Hollik:2011ps}. 

\begin{table}[t]
\begin{center}
\caption{Predicted asymmetries in the laboratory $A_{\rm lab}$
and the $t\bar t$ rest-frame $A_{t \bar t}$ at Tevatron. 
And relative amount of $u\bar u$, $d\bar d$ $gg$
initiated processes. Predictions are given also for 
samples with the top quark pair invariant mass $m_{t\bar t}$
above and below $450$~GeV. \label{tab:Attbar}}
\begin{tabular}{|c|lll|} \hline
laboratory     
          & $A_{\rm lab}$ & $m_{t\bar t}< 450$ GeV & $m_{t\bar t}> 450$ GeV \\ \hline 
QCD            & ~0.047 (7) & ~0.024 (2) & ~0.084 (9) \\
QED $u\bar u$  & ~0.0094    & ~0.0047    & ~0.0174    \\
QED $d\bar d$  & -0.0008    & -0.0004    & -0.0012    \\
weak $u\bar u$ & ~0.0011    & ~0.0006    & ~0.0021    \\
weak $d\bar d$ & -0.0003    & -0.0002    & -0.0005    \\ \hline
SM             & ~0.056 (7) & ~0.029 (2) & ~0.102 (9) \\ \hline
MCFM~\cite{Aaltonen:2011kc}         
               & ~0.038 (6)   &  & \\ \hline\hline
$t\bar t$ rest frame
          & $A_{t \bar t}$ & $m_{t\bar t}< 450$ GeV & $m_{t\bar t}> 450$ GeV \\ \hline 
QCD            & ~0.072 (9) & ~0.052 (4) & ~0.106 (11)\\
QED $u\bar u$  & ~0.0145    & ~0.0101    & ~0.0219    \\
QED $d\bar d$  & -0.0012    & -0.0010    & -0.0015    \\
weak $u\bar u$ & ~0.0018    & ~0.0012    & ~0.0027    \\
weak $d\bar d$ & -0.0005    & -0.0004    & -0.0006    \\ \hline
SM             & ~0.087 (10)& ~0.062 (4) & ~0.128 (11)\\ \hline
MCFM~\cite{Aaltonen:2011kc}         
               & ~0.058 (9)   & ~0.040 (6)   & ~0.088 (13)\\ \hline\hline
relative amount & inclusive & $m_{t\bar t}< 450$ GeV & $m_{t\bar t}> 450$ GeV \\ \hline 
$u \bar u$ & 0.78 & 0.76 & 0.82 \\
$d \bar d$ & 0.14 & 0.15 & 0.11 \\ 
$gg$       & 0.08 & 0.09 & 0.07 \\ \hline
\end{tabular}
\end{center}
\end{table}


In view of the strong dependence of the asymmetry on the invariant 
mass of the $t\bar t$ system, evident already from Figs.~\ref{fig:partonic}
and~\ref{fig:partonicshat}, we also study the dependence of the 
integrated asymmetries on a cut on the invariant mass of the 
$t\bar t$ system. Separating the events into two samples with 
$m_{t\bar t}$ smaller and larger than $450$~GeV respectively, 
the average value $A_{t\bar t} = 0.087 (10)$ moves down to 
$0.062 (4)$ and up to $0.128 (11)$ for the two choices. 
A similar behaviour is observed for the asymmetry $A_{\rm lab}$ defined 
in the laboratory frame.

It is interesting to compare these results with those based 
on a Monte Carlo prediction~\cite{Aaltonen:2011kc} 
based on MCFM~\cite{Campbell:1999ah}.
The enhancement factor of the SM result in Table~\ref{tab:Attbar}
compared to MCFM of about $1.5$ is easily understood: a 
factor $1.2$ originates from the inclusion of QED effects, that 
had been discussed in~\cite{Kuhn:1998jr} and improved 
in~\cite{Hollik:2011ps}. 
Another factor of about $1.3$ originates from normalizing with 
respect to the Born cross-section instead of the NLO result. 
Since the asymmetric part of the cross-section is presently 
known to LO only we consider the normalization to the LO
cross-section more 
plausible~\cite{Kuhn:1998kw,Kuhn:1998jr,Almeida:2008ug,Ahrens:2010zv}.

\begin{figure}[ht]
\begin{center}
\includegraphics[width=13cm]{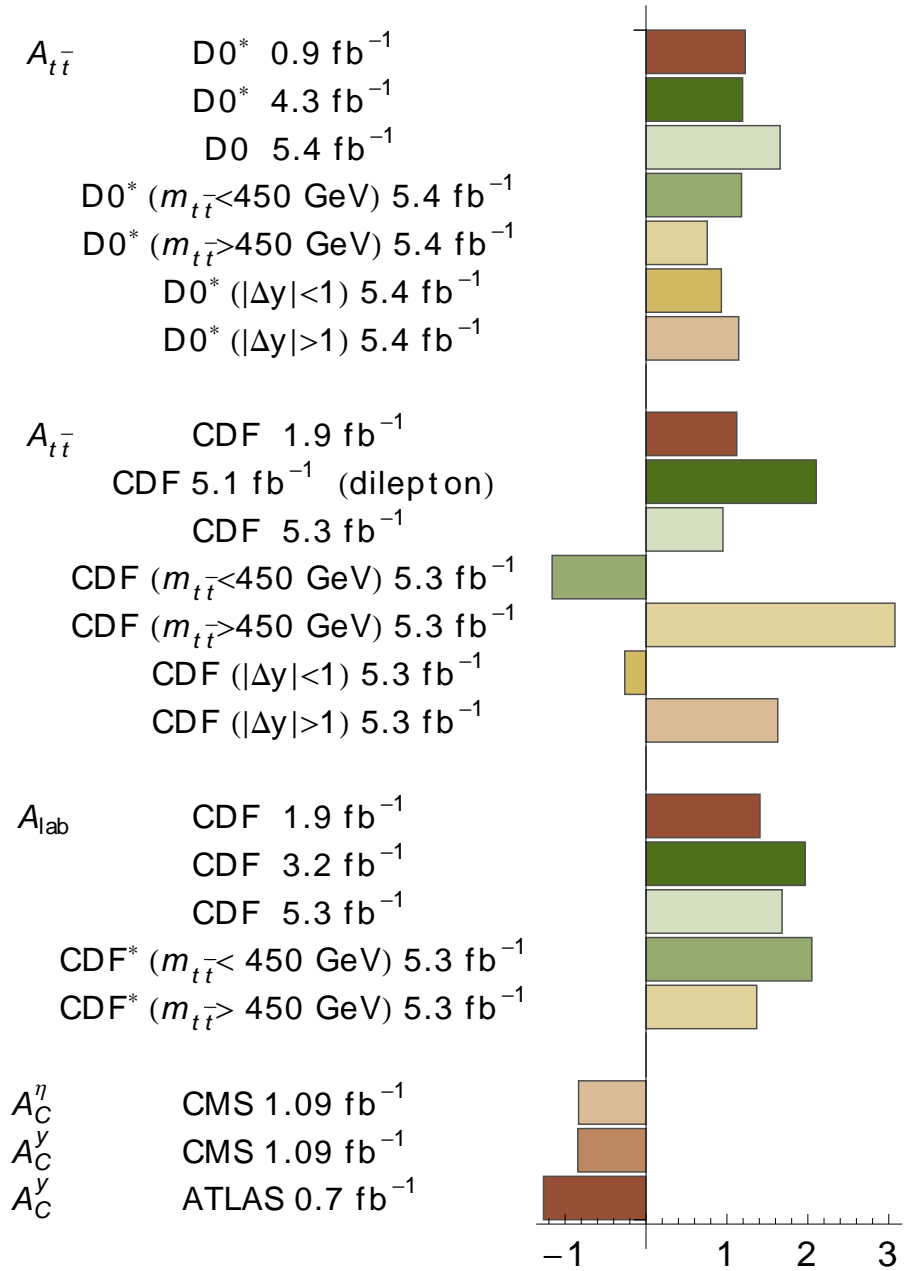}
\caption{Summary of experimental measurements of the charge asymmetry 
in comparison with the SM theoretical predictions. The histogram 
represents the pull of the discrepancy for each measurement.
The asterisk~$^*$ indicate ``reconstruction level'' results, the others 
are parton level. 
\label{fig:thexp}}
\end{center}
\end{figure}

For illustration, we compare these theoretical results 
in the SM with the most recent measurements at 
Tevatron~\cite{Abazov:2011rq,cdfcombined,cdfdilepton,Aaltonen:2011kc} 
as summarized in Table~\ref{tab:expTevatron}.
Predictions and results are shown both for $A_{\rm lab}$ and
$A_{t\bar t}$ and, when available, also split into two samples
with $m_{t\bar t}$ larger and smaller than $450$~GeV, and 
with $|\Delta y| = |y_t-y_{\bar t}|$ larger and smaller than $1$.
We note the nearly universal factor $\sim1.5$ between our result (SM) and the 
Monte Carlo simulation (MCFM / MC@NLO), which slightly softens the  
tension between theory and experiment. The errors from the choice 
of PDFs and factorization scale are small, the dominant uncertainty 
arises from varying the renormalization scale and hence the value 
of $\alpha_S$. If we would take the difference between LO and NLO 
prediction for the total production cross-section
as measure of the theory uncertainty, the error would increase 
up to $\pm 30\%$. A graphical illustration of the results in terms 
of the ''pull'' (measured in standard deviations) is shown in 
Fig.~\ref{fig:thexp} 
(errors from theory and experiment are combined quadratically.
For the results which refer to ''reconstruction
level'' we use the MCFM / MC@NLO results, multiplied by a factor $1.5$.)
The systematic upward shift of all but two Tevatron results is evident. 
The highest discrepancy, as has extensively been discussed in the 
literature, occurs for samples with $m_{t\bar t} > 450$~GeV and the charge 
asymmetry defined in the $t\bar t$ rest frame. 
Also shown in this Figure are preliminary results 
from CMS~\cite{CMS} and ATLAS~\cite{ATLAS} 
with a slight pull in the opposite direction.


\begin{table}[t]
\begin{center}
\caption{Recent experimental measurements 
of the asymmetry in the laboratory $A_{\rm lab}$ and in the 
$t\bar t$ rest frame $A_{t\bar t}$ at Tevatron. 
Results with $|\Delta y| = |y_t-y_{\bar t}|$ larger or smaller than $1$
are also summarized. Numbers with $^*$ refer to ``reconstruction 
level''~\cite{Aaltonen:2011kc,Abazov:2011rq}, 
the others to parton level. \label{tab:expTevatron}}
\begin{tabular}{|c|ccc|} \hline
laboratory
& $A_{\rm lab}$ & $m_{t\bar t}<450$~GeV & $m_{t\bar t}>450$~GeV \\ \hline 
SM (this work)             & 0.056 (7) & 0.029 (2) & 0.084 (9) \\
CDF~\cite{Aaltonen:2011kc} $5.3$~fb$^{-1}$ (l+jet)
& 0.150 $\pm$ 0.050 $\pm$ 0.024 
& 0.059 (34)$^*$ 
& 0.103 (49)$^*$  \\ 
MCFM/MC@NLO$^*$~\cite{Aaltonen:2011kc}         
               & 0.038 (6) & -0.008 (5)$^*$ & 0.022 (7)$^*$ \\ \hline\hline
$t\bar t$ rest frame
& $A_{t\bar t}$ & $m_{t\bar t}<450$~GeV & $m_{t\bar t}>450$~GeV \\ \hline 
SM (this work)             & 0.087 (10) & 0.062 (4) & 0.128 (11) \\
CDF~\cite{cdfdilepton} $5.1$~fb$^{-1}$ (dilep)
& 0.42 $\pm$ 0.15 $\pm$ 0.05 
& 
& 
\\
CDF~\cite{Aaltonen:2011kc} $5.3$~fb$^{-1}$ (l+jet)
& 0.158 $\pm$ 0.072 $\pm$ 0.017 
& -0.116 $\pm$ 0.146 $\pm$ 0.047 
& 0.475 $\pm$ 0.101 $\pm$ 0.049 
\\
CDF~\cite{cdfcombined} (combined)
& 0.20 $\pm$ 0.07 $\pm$ 0.02 
& 
& 
\\
MCFM~\cite{Aaltonen:2011kc}         
               & 0.058 (9)   & 0.040 (6)   & 0.088 (13)\\ 
D0~\cite{Abazov:2011rq} $5.4$~fb$^{-1}$ (l+jet) 
& 0.196 (65)
& 0.078 (48)$^*$
& 0.115 (60)$^*$ \\
MC@NLO~\cite{Abazov:2011rq}         
               & 0.050 (10)   & 0.013 (6)$^*$   & 0.043 (13)$^*$\\ 
\hline\hline
$t\bar t$ rest frame $|\Delta y|<1$
& $A_{t\bar t}$ & $m_{t\bar t}<450$~GeV & $m_{t\bar t}>450$~GeV \\ \hline 
SM (this work)             & 0.057 (4) & 0.053 (4) & 0.069 (5) \\ 
CDF~\cite{Aaltonen:2011kc} $5.3$~fb$^{-1}$ (l+jet)
& 0.026 $\pm$ 0.104 $\pm$ 0.056 & & \\
MCFM~\cite{Aaltonen:2011kc}         
               & ~0.039 (6)   &  & \\ 
D0~\cite{Abazov:2011rq} $5.4$~fb$^{-1}$ (l+jet) & 0.061 (41)$^*$ & & \\
MC@NLO~\cite{Abazov:2011rq}                    & 0.014 (6)$^*$ & & \\ 
\hline\hline
$t\bar t$ rest frame $|\Delta y|>1$
& $A_{t\bar t}$ & $m_{t\bar t}<450$~GeV & $m_{t\bar t}>450$~GeV \\ \hline 
SM (this work)             & 0.193 (15) & 0.149 (8) & 0.209 (15) \\ 
CDF~\cite{Aaltonen:2011kc} $5.3$~fb$^{-1}$ (l+jet)
& 0.611 $\pm$ 0.210 $\pm$ 0.147 & &   \\
MCFM~\cite{Aaltonen:2011kc}         
               & ~0.123 (18)  &  & \\ 
D0~\cite{Abazov:2011rq} $5.4$~fb$^{-1}$ (l+jet) & 0.213 (97)$^*$ & & \\
MC@NLO~\cite{Abazov:2011rq}                    & 0.063 (16)$^*$ & & \\ 
\hline
\end{tabular}
\end{center}
\end{table}


\begin{table}[t]
\begin{center}
\caption{SM asymmetries in the laboratory $A_{\rm lab}$ 
and the $t\bar t$ rest-frame $A_{t \bar t}$ for different 
cuts in $p_\perp^{t\bar t}$. \label{tab:AttbarpT}}

\begin{tabular}{|c|ccc|} \hline
laboratory     
          & $A_{\rm lab}$ & $m_{t\bar t}< 450$ GeV & $m_{t\bar t}> 450$ GeV \\ \hline 
$p_\perp^{t\bar t} < 10$ GeV            
               & ~0.090 (12)   & ~0.047 (3)   & ~0.161 (16)   \\ 
$p_\perp^{t\bar t} < 20$ GeV            
               & ~0.076 (10)   & ~0.040 (3)   & ~0.137 (13)   \\ \hline
$t\bar t$ rest frame
          & $A_{t \bar t}$ & $m_{t\bar t}< 450$ GeV & $m_{t\bar t}> 450$ GeV \\ \hline 
$p_\perp^{t\bar t} < 10$ GeV            
               & ~0.136 (16)   & ~0.097 (8)   & ~0.201 (19)   \\ 
$p_\perp^{t\bar t} < 20$ GeV            
               & ~0.115 (13)   & ~0.082 (7)   & ~0.171 (16)   \\ \hline
\end{tabular}
\end{center}
\end{table}


Let us now investigate the impact of cuts on hard gluon (and photon)
radiation on $A_{t\bar t} (Y)$. 
The dotted and dashed curves in Fig.~\ref{fig:AY}
show the effect of a cut on $p_\perp^{t\bar t}$ for values of 
$p_\perp^{\rm max} = 10$~GeV and $20$~GeV, respectively. 
An increase of the asymmetry by more than a factor 1.5 in the central 
region is observed for the most restrictive choice of $10$~GeV, 
and even a fairly loose $p_\perp^{\rm max} = 20$~GeV
modifies the asymmetry by up to a factor 1.3. 
The dependence on $Y$, the average rapidity of the $t$ and $\bar t$, 
is less flat than in the inclusive case. 
Similar enhancement factors are therefore 
also present in the integrated asymmetries, as displayed in 
Table~\ref{tab:AttbarpT}. A fairly similar behaviour is observed 
if cuts on both $m_{t\bar t}$ and $p_\perp^{t\bar t}$ are imposed, 
as shown in the third and fourth column of Table~\ref{tab:AttbarpT}. 
Note that the $p_\perp$ distribution of the $t\bar t$-system, 
as measured by the D0 Collaboration, seems to be at variance 
with the prediction based on the NLO Monte Carlo simulation~\cite{Demina}
(Leading order simulations like PYTHIA~\cite{Sjostrand:2006za} are by 
construction not able to correctly predict the asymmetry.)
Considering the fact that in the course of the experimental 
analysis $t\bar t$ events are separated 
into a sample with four jets (of $p_\perp$ larger than $20$~GeV)
and a sample with five and more jets (with different reconstruction 
algorithms and efficiencies) it seems important to verify that 
one nevertheless arrives at an unbiased inclusive sample.

\subsection{LHC}

\begin{figure}[htb]
\begin{center}

\begin{picture}(200,200)(0,0)
\SetWidth{1.1}
\SetScale{1.}

\SetOffset(10,170)

\ArrowLine(-60,0)(-33,0)
\ArrowLine(50,0)(-27,0)
\ArrowLine(-30,0)(-100,30)
\ArrowLine(-30,0)(-10,-30)
\Text(0,-50)[]{preferred}
\Text(-60,-15)[]{$\bar q$}
\Text(40,-15)[]{$q$}
\Text(-100,15)[]{$Q$}
\Text(0,-30)[]{$\bar Q$}

\SetOffset(150,170)

\ArrowLine(60,0)(33,0)
\ArrowLine(-50,0)(27,0)
\ArrowLine(30,0)(100,30)
\ArrowLine(30,0)(10,-30)
\Text(0,-50)[]{preferred}
\Text(60,-15)[]{$\bar q$}
\Text(-40,-15)[]{$q$}
\Text(100,15)[]{$Q$}
\Text(0,-30)[]{$\bar Q$}

\SetOffset(10,60)

\ArrowLine(-60,0)(-33,0)
\ArrowLine(50,0)(-27,0)
\ArrowLine(-30,0)(-100,30)
\ArrowLine(-30,0)(-10,-30)
\Text(0,-50)[]{suppressed}
\Text(-60,-15)[]{$\bar q$}
\Text(40,-15)[]{$q$}
\Text(-100,15)[]{$\bar Q$}
\Text(0,-30)[]{$Q$}

\SetOffset(150,60)

\ArrowLine(60,0)(33,0)
\ArrowLine(-50,0)(27,0)
\ArrowLine(30,0)(100,30)
\ArrowLine(30,0)(10,-30)
\Text(0,-50)[]{suppressed}
\Text(60,-15)[]{$\bar q$}
\Text(-40,-15)[]{$q$}
\Text(100,15)[]{$\bar Q$}
\Text(0,-30)[]{$Q$}

\end{picture}
\caption{Preferred and suppressed configurations at the
LHC.\label{fig:preferred}}
\end{center}
\end{figure}
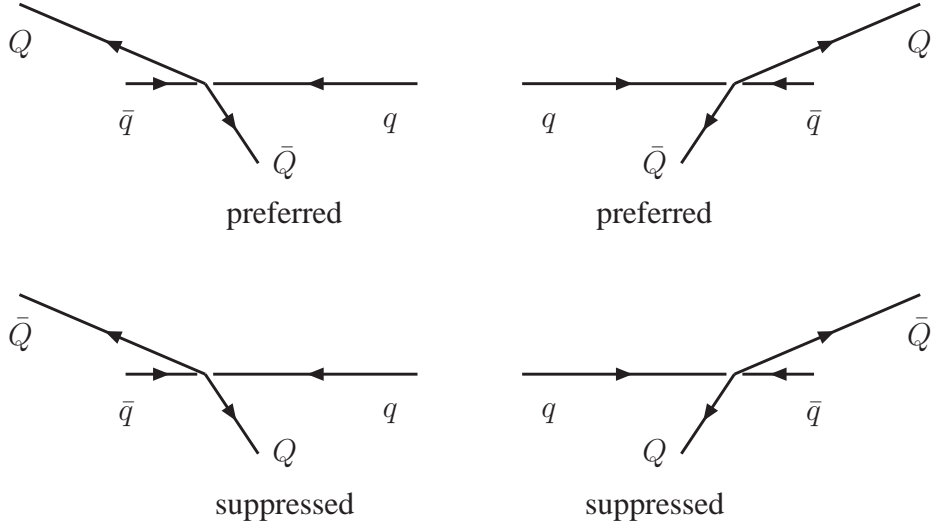

\begin{figure}[ht]
\begin{center}
\includegraphics[width=8cm]{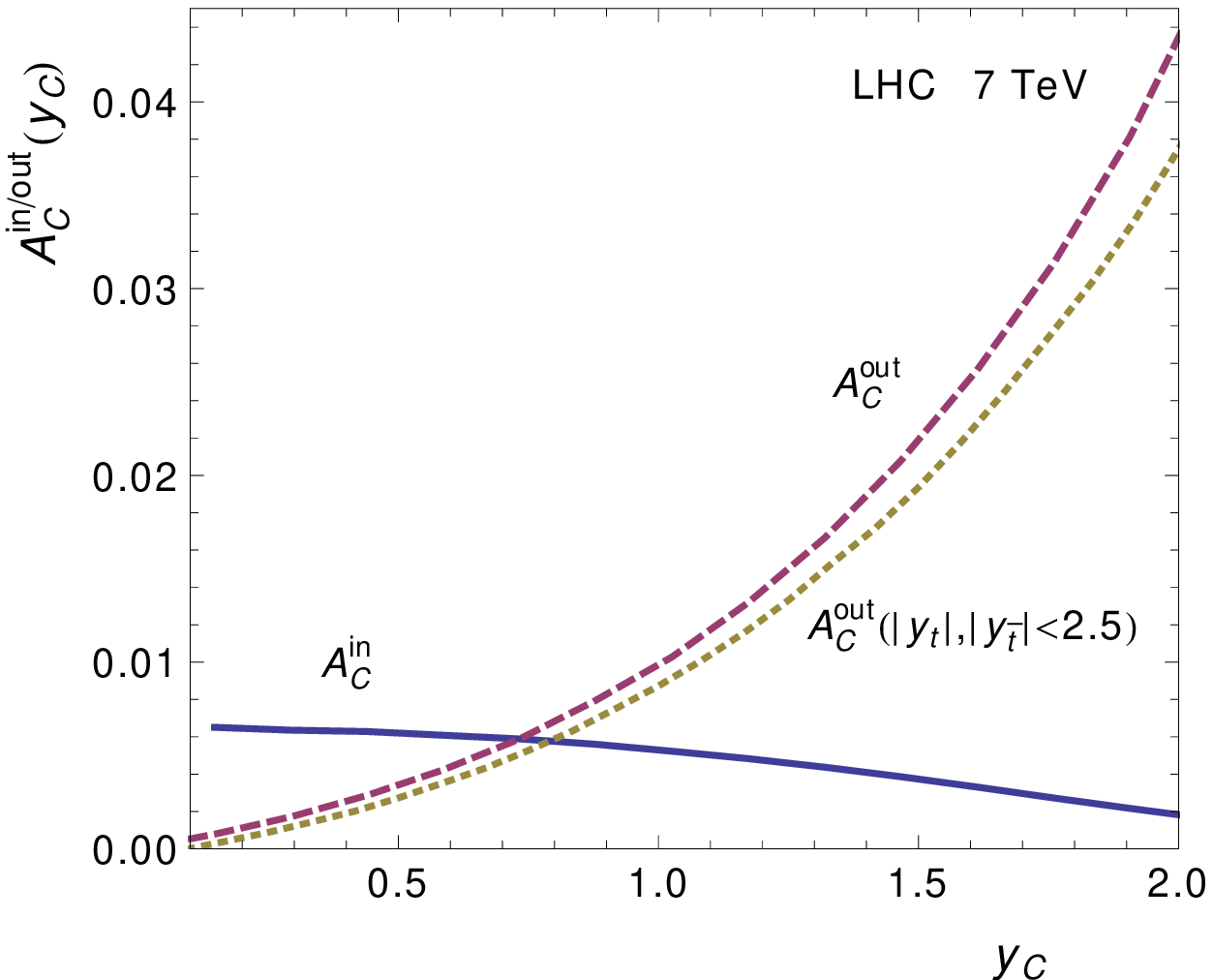}
\includegraphics[width=8cm]{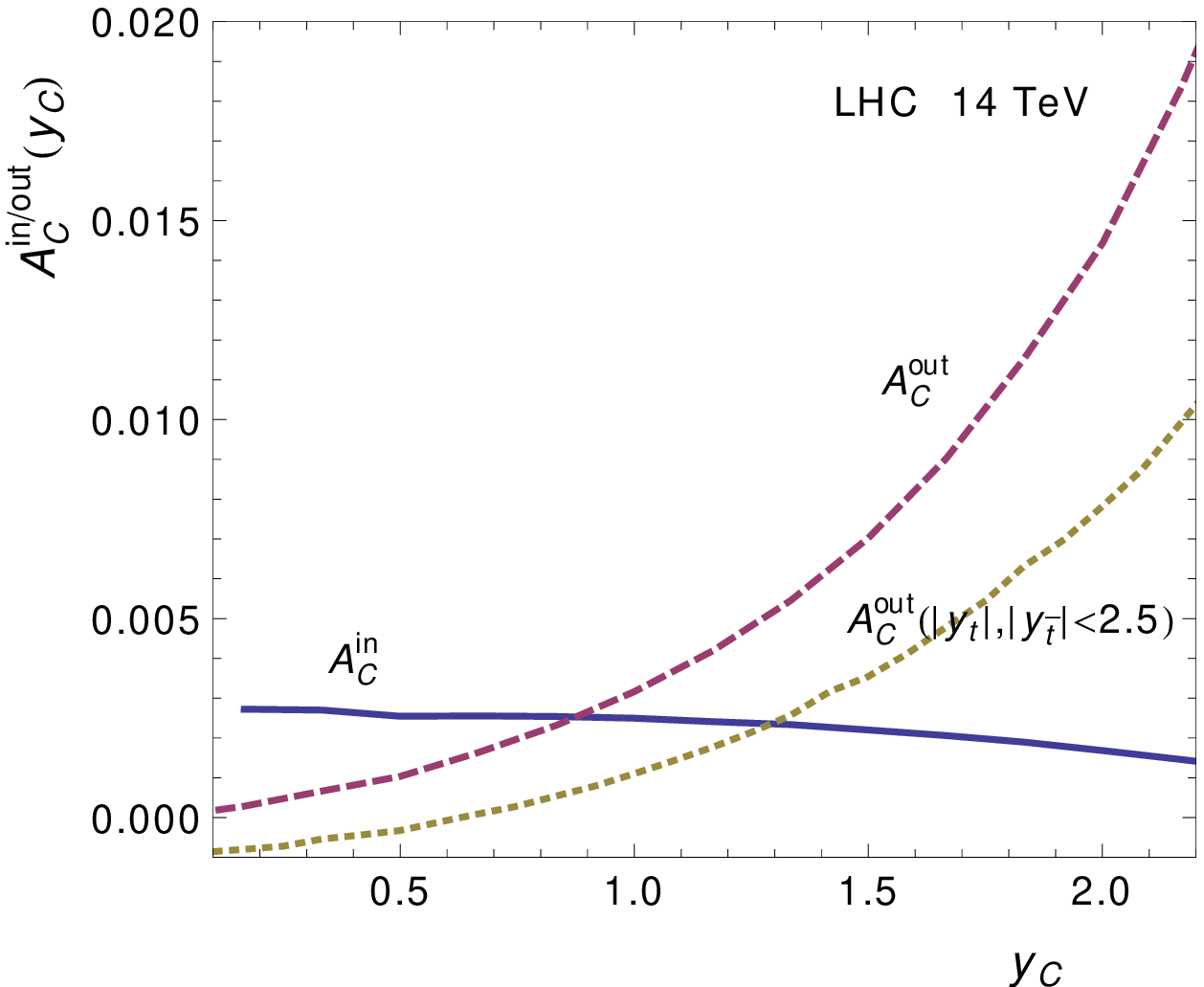}
\caption{In and out charge asymmetries at the LHC with 
$\sqrt{s}=7$ TeV and $14$ TeV as a function of the 
cut $y_C$ in rapidity. The out charge asymmetry 
$A_C^{\rm out}(y_C)$ is also calculated with an upper cut
in the rapidity of the top and antitop quarks 
$|y_t|,|y_{\bar t}|<2.5$. \label{fig:inout}}
\end{center}
\end{figure}

As discussed in~\cite{Antunano:2007da,Kuhn:1998kw,Kuhn:1998jr}
it is possible to investigate the charge asymmetry also in proton-proton
collisions at the LHC, exploiting the small $t\bar t$ sample 
produced in annihilation of valence quarks and antiquarks from 
the sea. As illustrated in Fig.~\ref{fig:preferred}
production of quarks with larger rapidities will be preferred, 
antitop quarks will be produced more frequently at smaller 
rapidities. This observation suggests to define 
the cut-dependent asymmetries 
\beq
A_C^{\rm in}(y_C) = \frac{N(|y_{\bar t}|\le y_C)-N(|y_t|\le y_C)}
{N(|y_t|\le y_C)+N(|y_{\bar t}|\le y_C)}
\label{eq:aCin}
\eeq
and
\beq
A_C^{\rm out}(y_C) = \frac{N(|y_t|>y_C)-N(|y_{\bar t}|>y_C)}
{N(|y_t|>y_C)+N(|y_{\bar t}|>y_C)}~,
\eeq
which serve to characterize the depletion of top quarks in the 
central region ($A_C^{\rm in}(y_C)>A_C^{\rm out}(y_C)$ for $y_C<0.7$ 
approximately~\cite{Antunano:2007da}), and 
their enhancement at larger rapidities 
($A_C^{\rm in}(y_C) < A_C^{\rm out}(y_C)$ for $y_C>0.7$). 
Note that we have defined $A_C^{\rm in}(y_C)$ in~\Eq{eq:aCin} 
with the opposite sign than 
in Ref.~\cite{Antunano:2007da,Ferrario:2008wm,Ferrario:2009ee}
such that both $A_C^{\rm in}(y_C)$ and $A_C^{\rm out}(y_C)$ 
are positive in the SM.
The dependence of $A_C^{\rm in}$ and $A_C^{\rm out}$ on $y_C$
is shown in Fig.~\ref{fig:inout} 
for $\sqrt{s}=7$~TeV (left plot) and $14$~TeV (right plot). 
As one can observe in these Figures, $A_C^{\rm out}$ is much larger 
than $A_C^{\rm in}$ at large values of the rapidity cut $y_C$. 
This is because the central region is dominated by gluon fusion 
processes, while the sample with large rapidities has a larger 
relative content of $q\bar q$ initiated events. 
The statistical significance of both observables is, however, 
very similar~\cite{Hewett:2011wz} because the larger size of the asymmetry 
$A_C^{\rm out}$ with respect to $A_C^{\rm in}$
is compensated by the lower rate of events at larger rapidities.
We consider also the cut-independent charge asymmetries 
\beq
A_C^\eta = \frac{N(\Delta_\eta>0)-N(\Delta_\eta<0)}
{N(\Delta_\eta>0)+N(\Delta_\eta<0)}
\eeq
and
\beq
A_C^y = \frac{N(\Delta_y>0)-N(\Delta_y<0)}
{N(\Delta_y>0)+N(\Delta_y<0)}~,
\eeq
where $\Delta_\eta = |\eta_t|-|\eta_{\bar t}|$
and $\Delta_y = |y_t|-|y_{\bar t}|$, which have 
been used in the recent CMS~\cite{CMS} and ALTAS~\cite{ATLAS} analysis.
The SM predictions for the integrated asymmetries are 
listed Table~\ref{tab:AttbarmultiTeV} for different center-of-mass 
energies of the LHC, together with the experimental results for 
$\sqrt{s}=7$~TeV. Both experiments obtain negative asymmetries, 
although compatible with the SM prediction within uncertainties.


\begin{table}[t]
\begin{center}
\caption{SM cut-independent asymmetries $A_\eta$  and $A_y$ 
at different LHC energies. 
\label{tab:AttbarmultiTeV}}

\begin{tabular}{|l|cc|} \hline     
               & $A_C^\eta$   & $A_C^y$      \\ \hline 
LHC 7 TeV      & 0.0136 (8)   & 0.0115 (6)    \\
LHC 8 TeV      & 0.0122 (7)   & 0.0102 (5)    \\
LHC 10 TeV     & 0.0101 (6)   & 0.0082 (4)    \\
LHC 12 TeV     & 0.0087 (5)   & 0.0068 (3)    \\
LHC 14 TeV     & 0.0077 (4)   & 0.0059 (3)    \\\hline\hline
LHC 7 TeV CMS~\cite{CMS} $1.09$~fb$^{-1}$    
 & -0.016 $\pm$ 0.030 ${}^{+0.010}_{-0.019}$ 
 & -0.013 $\pm$ 0.026 ${}^{+0.026}_{-0.021}$  \\
LHC 7 TeV ATLAS~\cite{ATLAS} $0.7$~fb$^{-1}$    &   
 & -0.024 $\pm$ 0.016 $\pm$ 0.023   \\\hline
\end{tabular}
\end{center}
\end{table}

\begin{figure}[ht]
\begin{center}
\includegraphics[width=8cm]{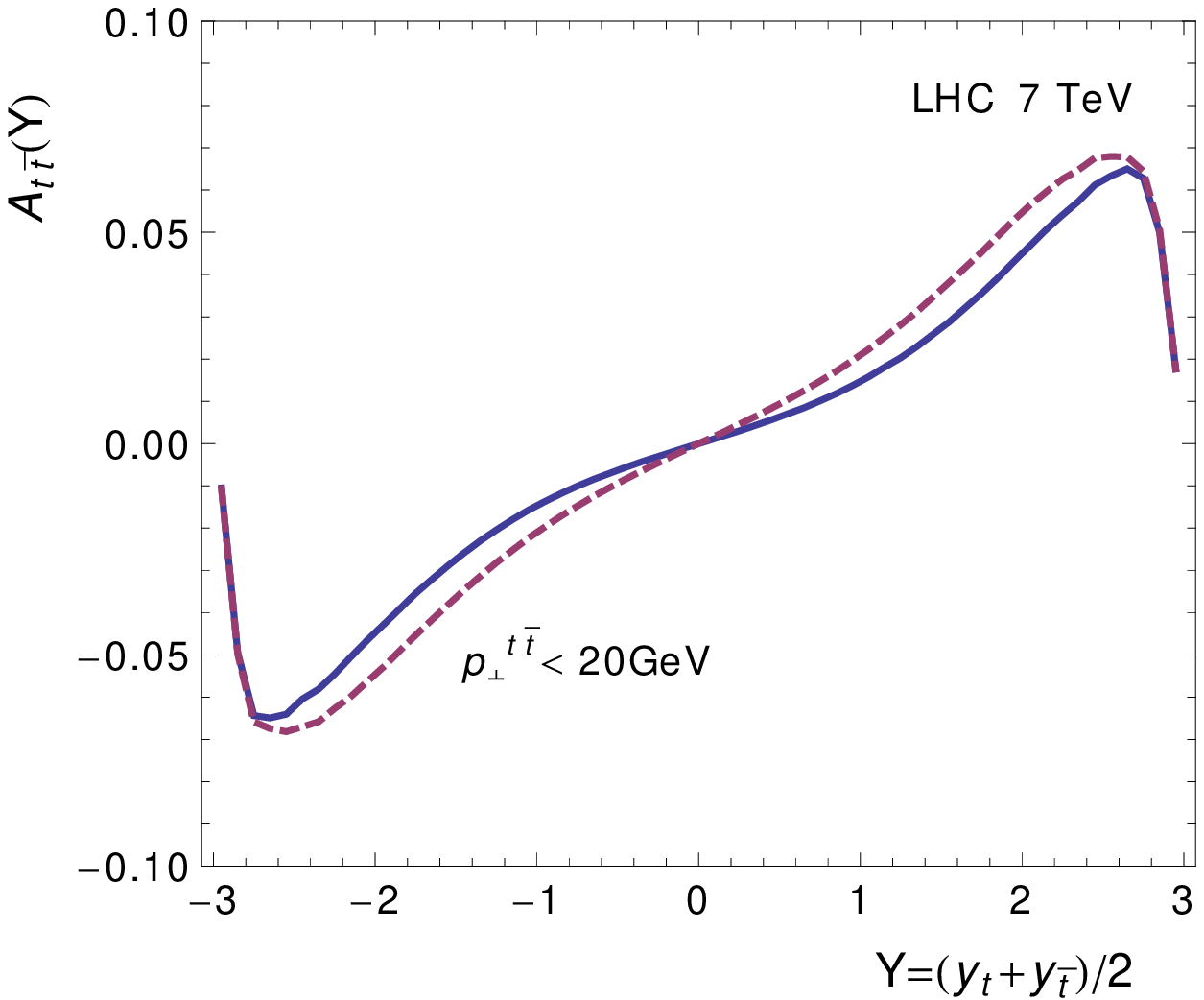}
\includegraphics[width=8cm]{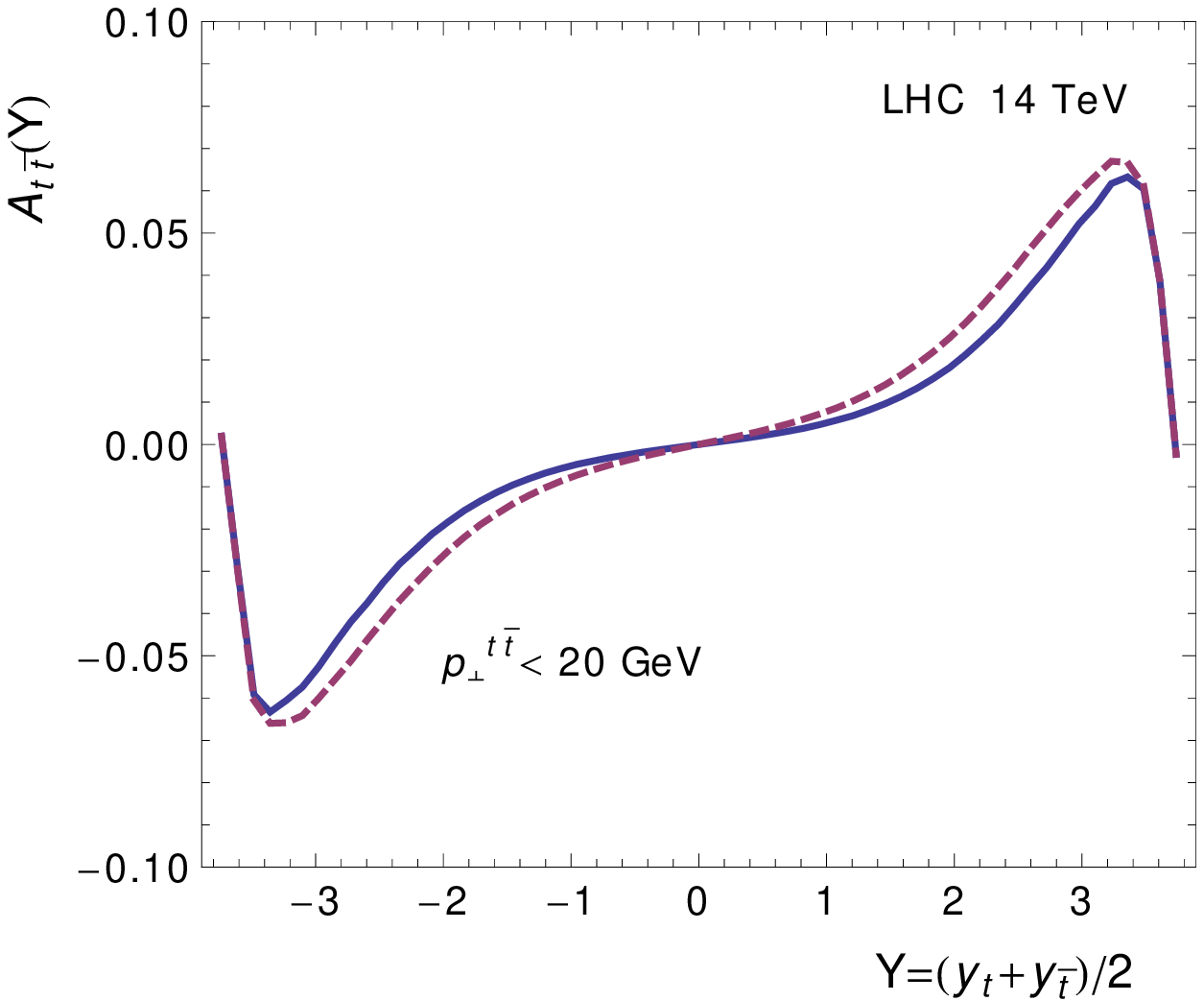}
\caption{Pair charge asymmetry at the LHC with $\sqrt{s}=7$~TeV 
and $14$~TeV as a function of the mean rapidity $Y=(y_t+y_{\bar t})/2$. 
Solid line: no cut on $p_\perp^{t\bar t}$, dashed line: $p_\perp^{\rm max} = 20$~GeV.
\label{fig:pair}}
\end{center}
\end{figure}
\begin{figure}[ht]
\begin{center}
\includegraphics[width=8cm]{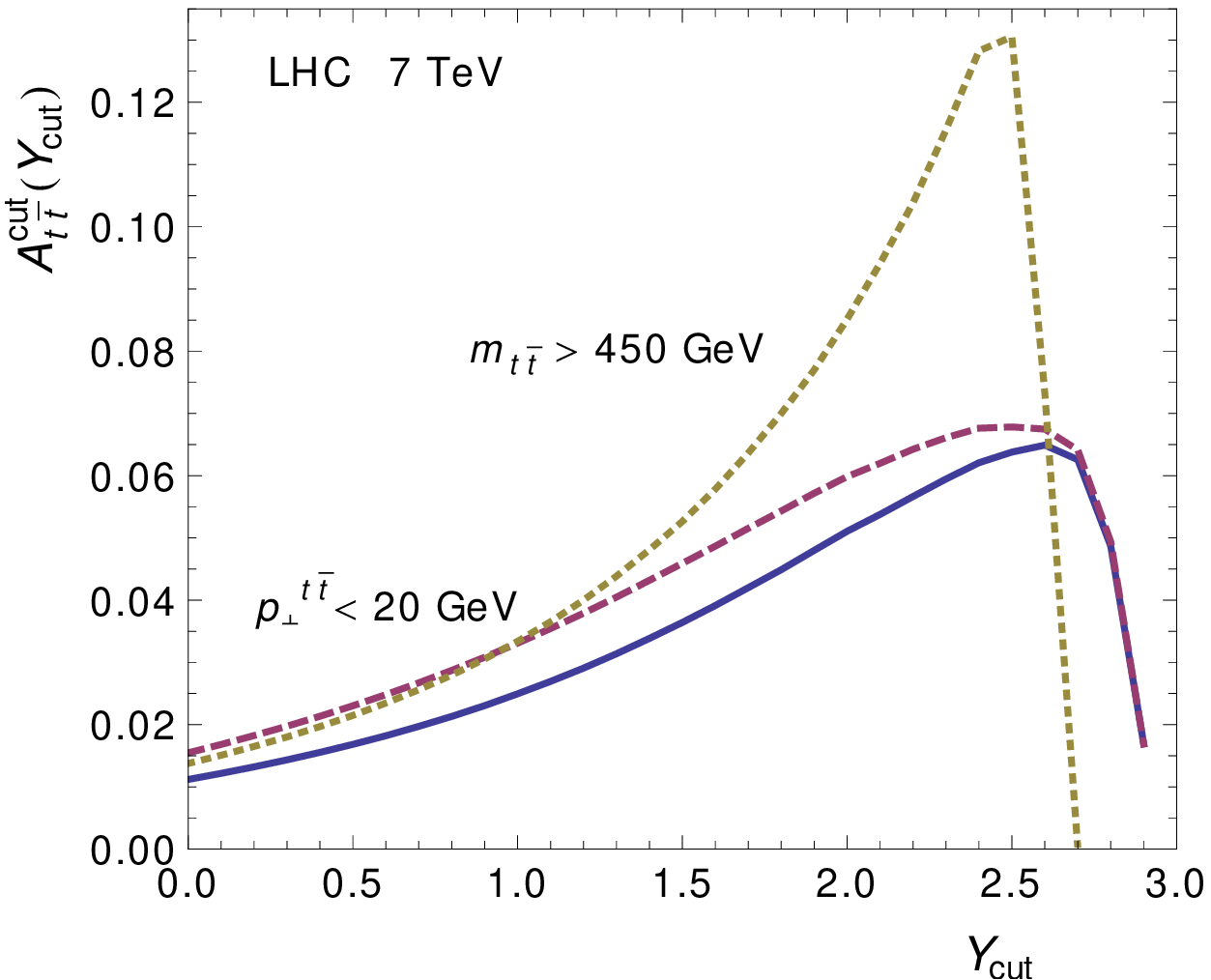}
\includegraphics[width=8cm]{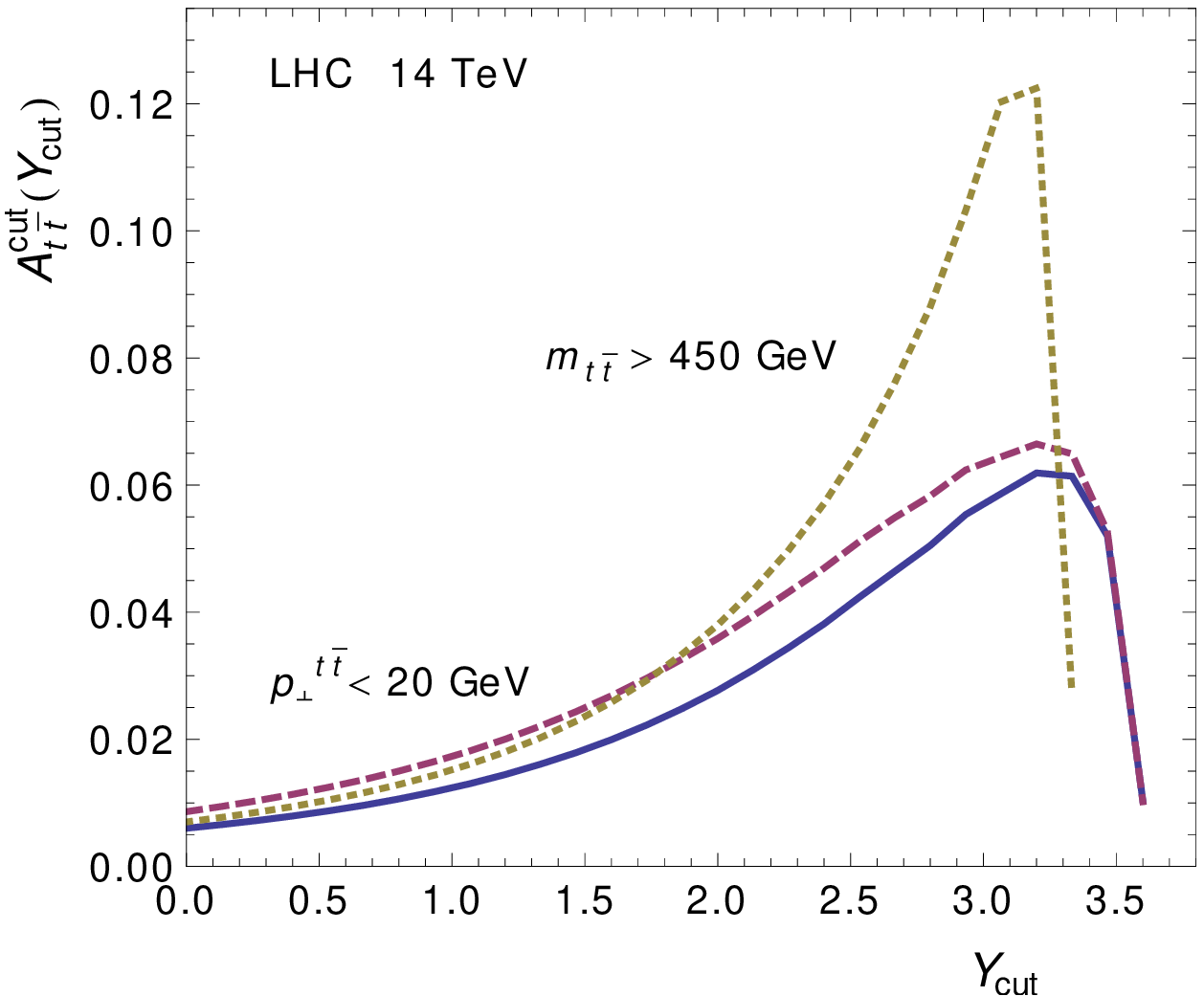}
\includegraphics[width=8cm]{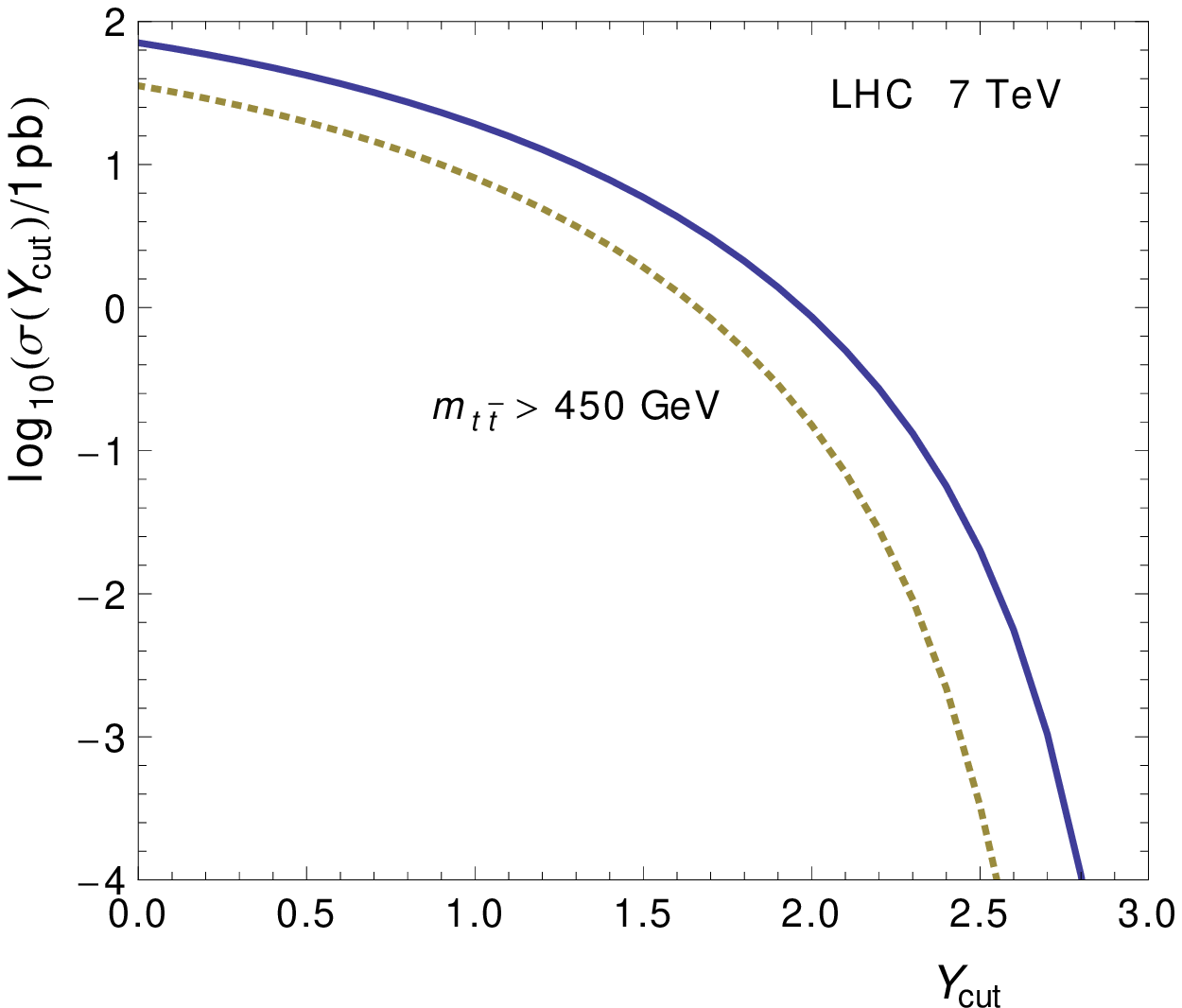}
\includegraphics[width=8cm]{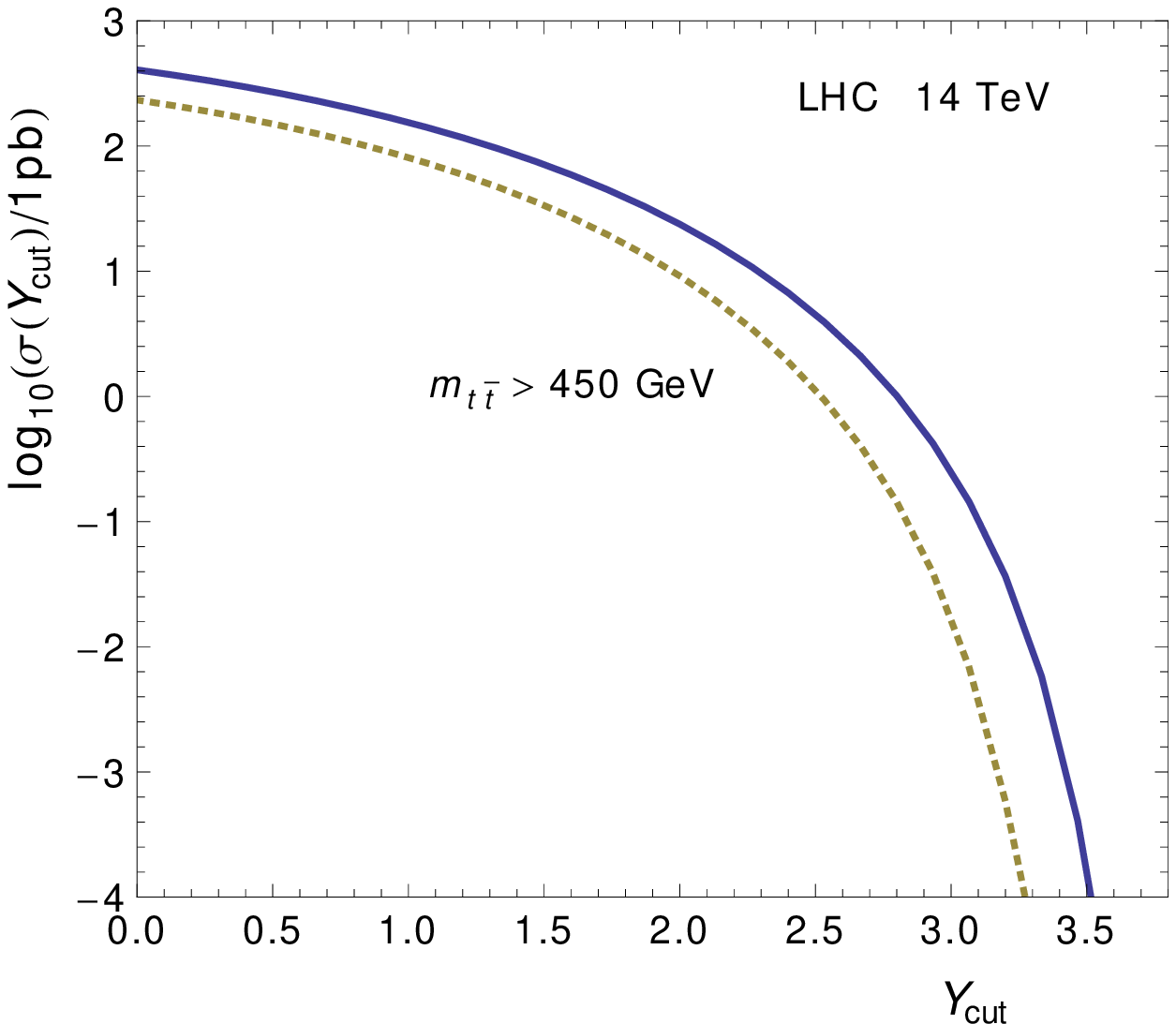}
\caption{Integrated pair charge asymmetry and integrated cross-section
as a function of $Y_{\rm cut}$ at the LHC with $\sqrt{s}=7$~TeV and $14$~TeV. 
Solid line: without extra cuts, dashed line: $p_\perp^{\rm max} = 20$~GeV, 
dotted line: $m_{t\bar t}>450$~GeV. \label{fig:paircut}}
\end{center}
\end{figure}

Top quark production in proton-proton collisions is dominated 
by gluon fusion, which, in turn, is dominant in the central region. 
Conversly, quark-antiquark annihilation will be more enriched
for events with $t\bar t$ at larger rapidities (and larger 
$m_{t\bar t}$). 
This suggest to employ the definition of~\Eq{eq:pair}, which 
is essentially the asymmetry in the $t\bar t$ rest frame, also 
for the present case, and concentrate on $t\bar t$ events at 
large rapidities. The prediction for $A_{t\bar t} (Y)$ is shown 
in Fig.~\ref{fig:pair}
for $\sqrt{s}=7$~TeV (left plot) and $14$~TeV (right plot).
Note that $A_{t\bar t} (Y)$ is now, by construction, 
an antisymmetric function of $Y$.
Since most of the charge asymmetry is concentrated at large rapidities 
the statistical significance of any measurement will be enhanced, 
if the sample is restricted to larger rapidities.
Let us therefore define the quantity
\beq
A_{t\bar t}^{\rm cut} \, (Y_{\rm cut})=\frac{N(y_t>y_{\bar t})-N(y_{\bar t}>y_t)}
{N(y_t>y_{\bar t})+N(y_{\bar t}>y_t)}~,
\label{eq:paircut}
\eeq
where $(y_t+y_{\bar t})/2>Y_{\rm cut}$~\footnote{The asymmetry  
$A_{t\bar t}^{\rm cut} \, (Y_{\rm cut})$ of the sample with 
$(y_t+y_{\bar t})/2<-Y_{\rm cut}$ is of the same magnitude but of 
opposite sign. Both samples can be combined to enhance the statistical 
significance of the measurement by introducing a flip of sign in the
definition of $A_{t\bar t}^{\rm cut} \, (Y_{\rm cut})$ for events 
with $(y_t+y_{\bar t})/2<-Y_{\rm cut}$.}.
The prediction for $A^{\rm cut}_{t\bar t}(Y_{\rm cut})$ 
is shown in Fig.~\ref{fig:paircut}
for $\sqrt{s}=7$~TeV (left upper plot) and $14$~TeV (right upper plot).
Note that the prediction includes QED and weak corrections, 
which amount to roughly a factor 1.1. 
To estimate the statistical sensitivity of such a measurement, 
the cross section for events with $t\bar t$ at large rapidities 
$|(y_t+y_{\bar t})/2|>Y_{\rm cut}$ is shown in Fig.~\ref{fig:paircut}
(lower plots) as a function of $Y_{\rm cut}$. 
Using $\sqrt{s}=7$ and an integrated 
luminosity of $20$~fb$^{-1}$ as example, more than $4\times 10^5$ 
events are expected for $Y_{\rm cut} = 1.$
Even allowing for a significant reduction of the sample by small efficiencies, 
the asymmetry $A_{t\bar t}^{\rm cut}(Y_{\rm cut})$ of more than $2\%$ 
could be observed by experiment.

As discussed before, cuts on $p_{\perp}^{t\bar t}$ and $m_{t\bar t}$
may lead to a significant modification of the asymmetry. 
The impact of a cut $p_{\perp}^{t\bar t}<20$~GeV 
on $A_{t\bar t}\, (Y)$ and $A_{t\bar t}^{\rm cut} \, (Y_{\rm cut})$ 
is shown Figs.~\ref{fig:pair} and \ref{fig:paircut} (upper plots)  
as dashed curves. Numerical results for the integrated pair charge asymmetry
with $Y_{\rm cut} = 0.7$ are listed in Table~\ref{tab:AttbarYcutmultiTeV}
for the inclusive sample, $A_{t\bar t}^{\rm cut}(Y_{\rm cut}=0.7)$,
and for subsamples with $m_{t\bar t}$ larger and smaller than $450$~GeV.
By definition $A_{t\bar t}^{\rm cut}(Y_{\rm cut}=0) = A_C^y$, 
for which results are listed in Table~\ref{tab:AttbarmultiTeV}.

\begin{table}[htb]
\begin{center}
\caption{SM integrated pair charge asymmetry at different LHC energies 
for $Y_{\rm cut}=0.7$, for the inclusive sample, $A_{t\bar t}^{cut}(Y_{\rm cut}=0.7)$, 
and for subsamples with $m_{t\bar t}$ larger and smaller than $450$~GeV.
\label{tab:AttbarYcutmultiTeV}}
\begin{tabular}{|l|ccc|} \hline     
               & $A_{t\bar t}^{\rm cut}(Y_{\rm cut}=0.7)$ & $m_{t\bar t}< 450$~GeV  
& $m_{t\bar t}> 450$~GeV   \\ \hline 
LHC 7 TeV      & 0.0203 (8)   & 0.0148 (5)   & 0.0263 (8) \\
LHC 8 TeV      & 0.0178 (6)   & 0.0128 (4)   & 0.0224 (7) \\
LHC 10 TeV     & 0.0142 (5)   & 0.0104 (4)   & 0.0174 (5) \\
LHC 12 TeV     & 0.0117 (4)   & 0.0085 (3)   & 0.0143 (4) \\
LHC 14 TeV     & 0.0100 (4)   & 0.0075 (3)   & 0.0121 (4) \\\hline
\end{tabular}
\end{center}
\end{table}


\section{Charge asymmetry beyond the SM at the LHC}

As noted in~\cite{Antunano:2007da}, the asymmetry induced by a 
``conventional'' axigluon $G$, i.e. with identical axial-vector coupling 
$g_S$ for all quarks and assuming $m_G>2m_t$, is negative. This 
has lead to stringent bounds on $m_G$~\cite{Ferrario:2009bz}.
On the other hand the apparent positive excess as observed at the Tevatron 
has lead to numerous suggestions for physics beyond the SM
which, however, are difficult to reconcile with other experimental 
facts (For a recent discussion see e.g. \cite{Westhoff}).
On the contrary,  both measurements at CMS~\cite{CMS} and 
ATLAS~\cite{ATLAS} point towards negative asymmetries, 
although still with large uncertainties and compatible with the SM. 
Independently of these theoretical and experimental considerations 
it will be interesting to investigate the same phenomena at the LHC. 
As discussed previously in~\cite{Kuhn:1998kw} (see e.g. Figs. 11 and 12) 
it is possible to identify kinematic regions where $q\bar q$ annihilation 
into $t\bar t$ is comparable or even larger than gluon fusion, 
and the charge asymmetry is suitable to probe new physics beyond the SM
in that kinematical regions. 
We do not try in this section to focus on a particular model 
that would fit better the Tevatron anomaly and then extrapolate 
that model to the LHC. Rather, we try to investigate the 
power of the charge asymmetry defined in \Eq{eq:paircut} to discriminate 
among two of the simplest models giving rise to a BSM charge asymmetry, 
one of them giving a positive excess, and another one leading to 
a negative contribution. 

The most recent measurements at the LHC of the dijet 
cross-section~\cite{Aad:2011fq,Chatrchyan:2011ns} 
impose stringent constraints on axigluon masses below 3 TeV. 
Still, those limits can be relaxed when considering top quark 
pair production in models in which the coupling of the axigluon 
vector boson to light quarks is much smaller than the coupling
to the top quark. As the simplest case, however, we consider 
here the case were the extra gauge boson couples with the 
same strength to light and top quarks, in two different scenarios: 
a flavour universal case (octet U), i.e. vector-axial couplings to 
light and top quarks are equal to the strong coupling $g_S$ multiplied by 
a factor $1.8$, $g_A^q=g_A^t=1.8$, 
and a flavour non-universal case (octet A) with vector-axial couplings
to light and top quarks of the opposite sign~\cite{Ferrario:2009bz},
$g_A^q=-g_A^t=1.8$.   
The latter naturally produces positive contributions to the 
charge asymmetry, and has been advocated as one of the possible 
solutions to the Tevatron anomaly. The former gives a negative 
contribution to the charge asymmetry, and is disfavored 
by most of the Tevatron measurements, but still compatible 
with some of the measurements within 2$\sigma$ (see Fig.~\ref{fig:thexp}).
In both cases we consider an axigluon mass of $3$~TeV as benchmark model. 

\begin{figure}[ht]
\begin{center}
\includegraphics[width=8cm]{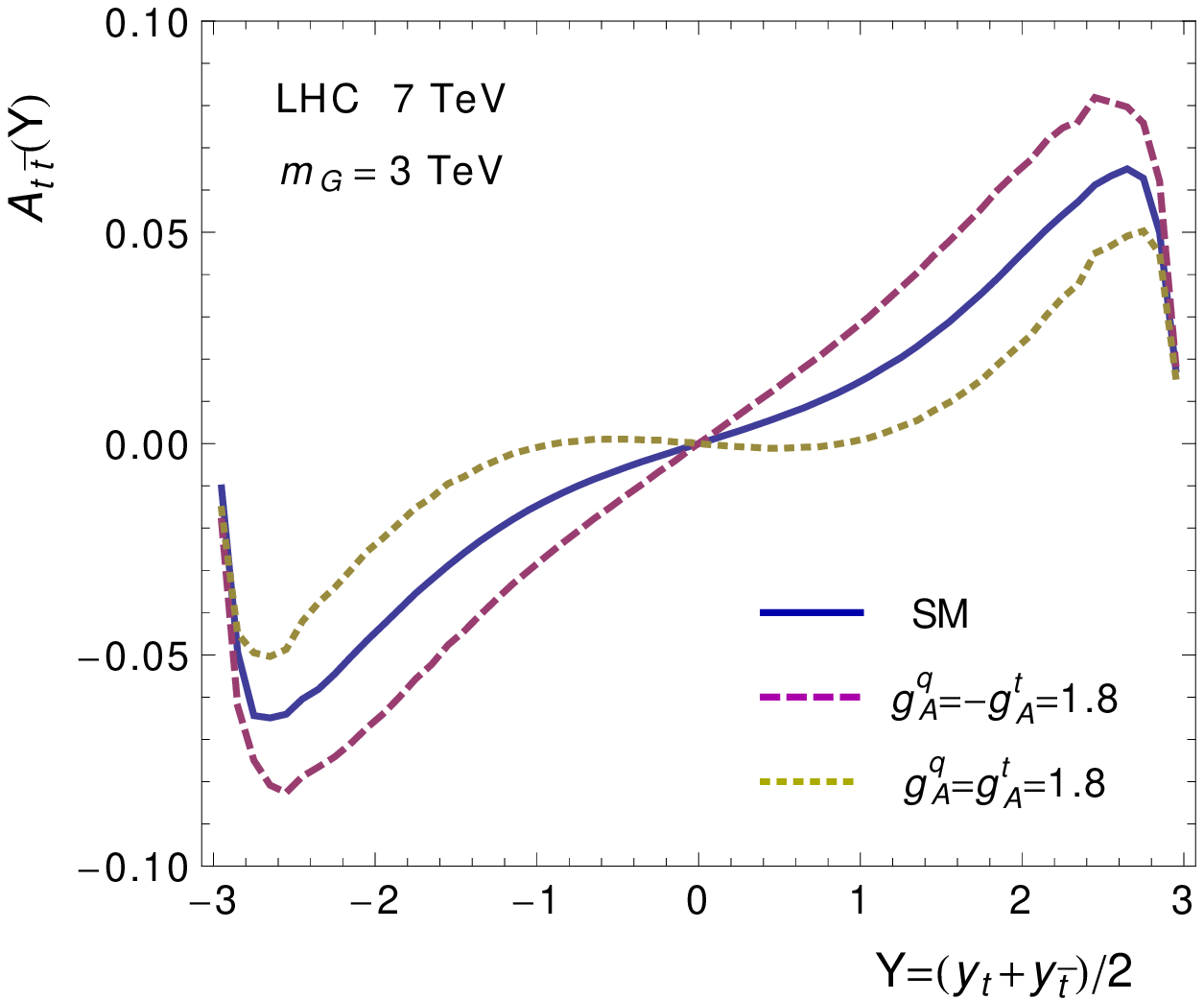}
\includegraphics[width=8cm]{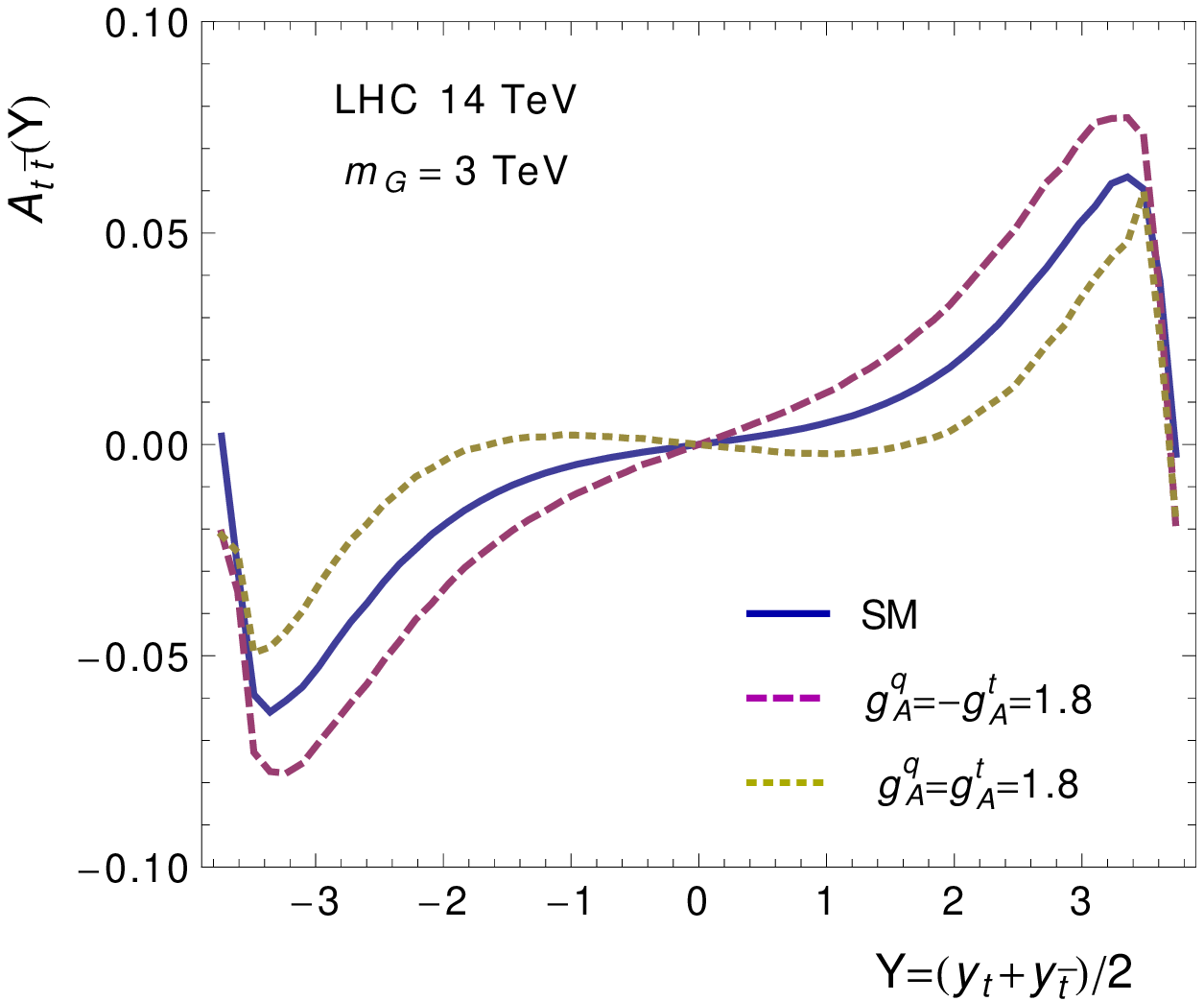}
\includegraphics[width=8cm]{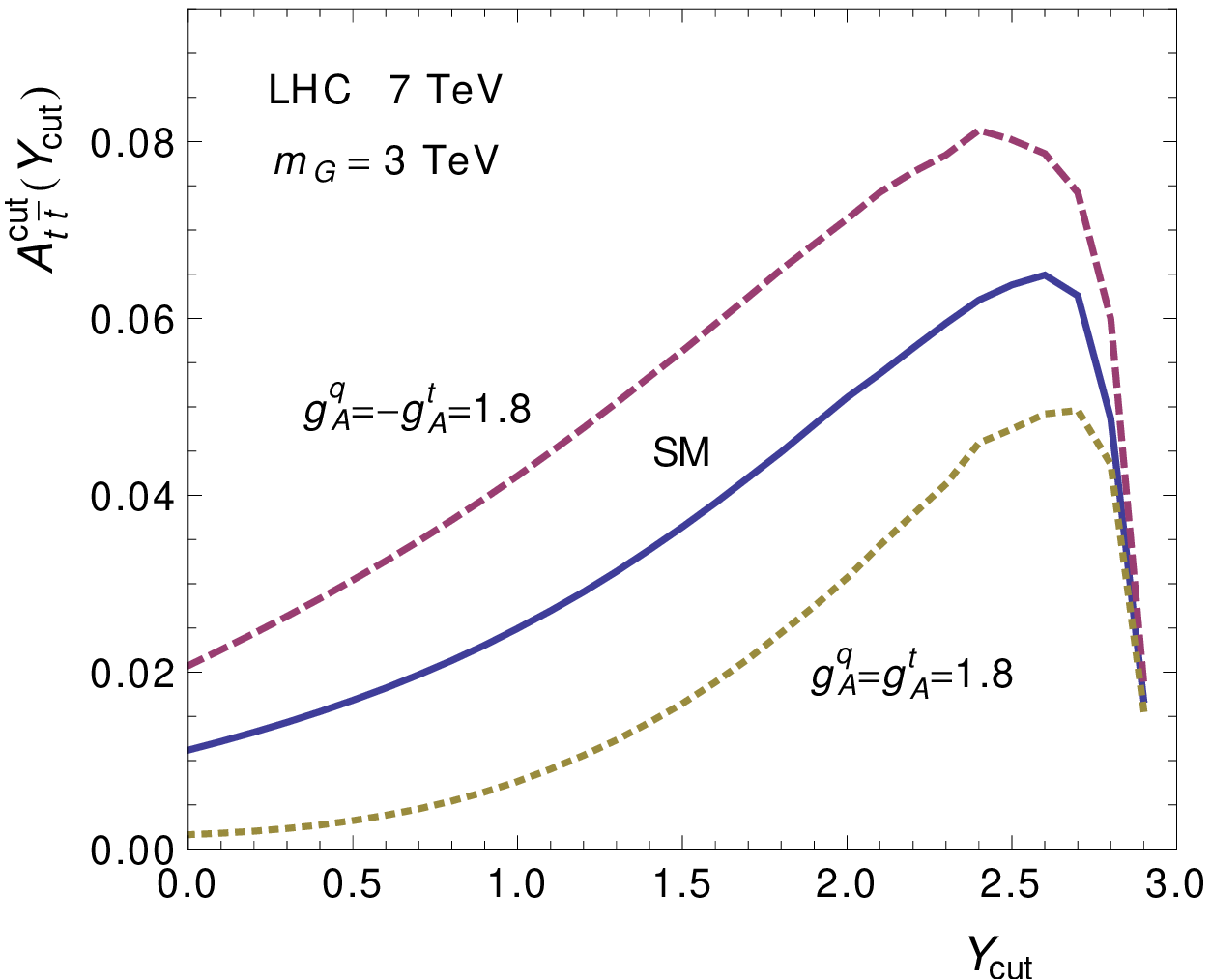}
\includegraphics[width=8cm]{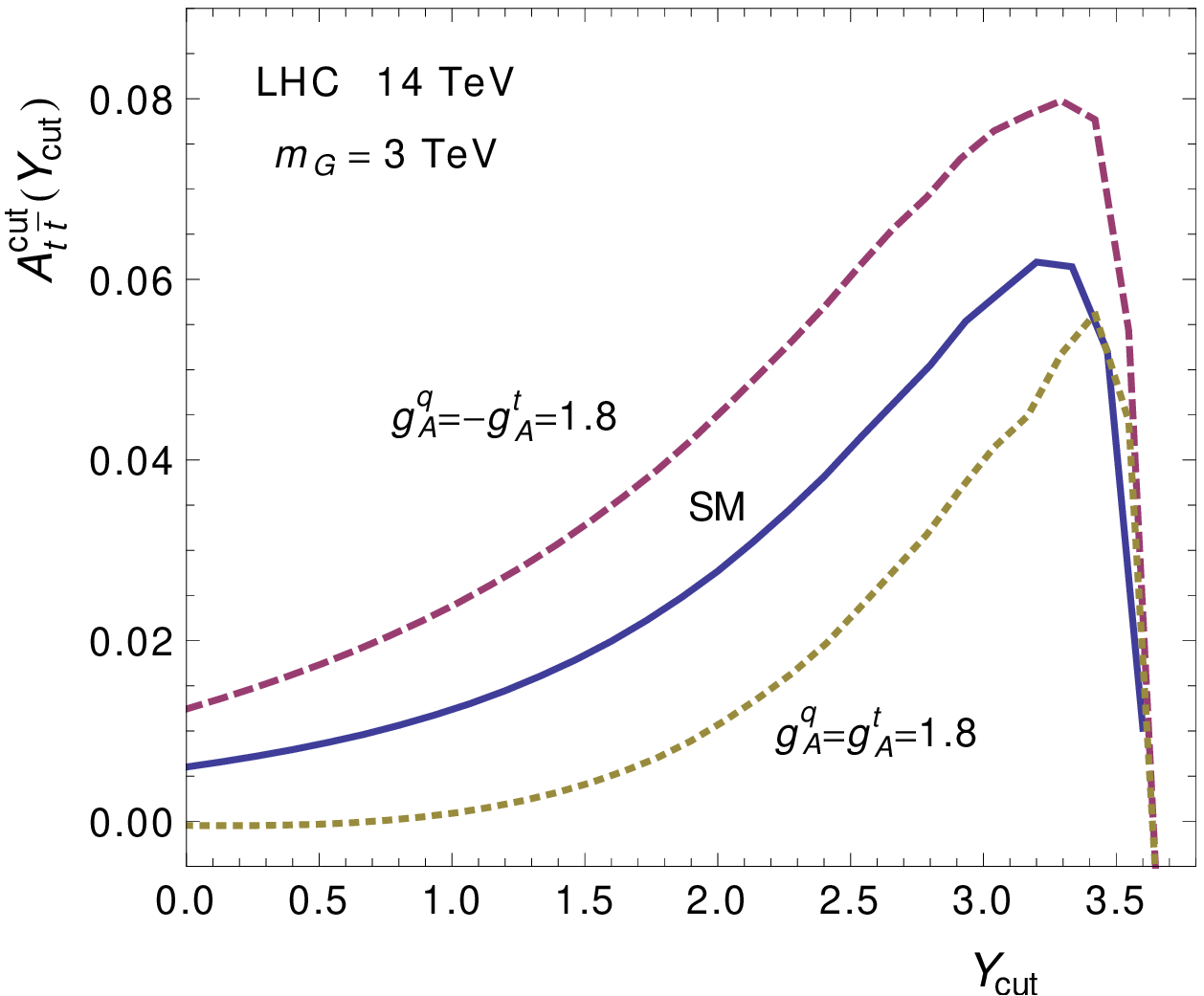}
\includegraphics[width=8cm]{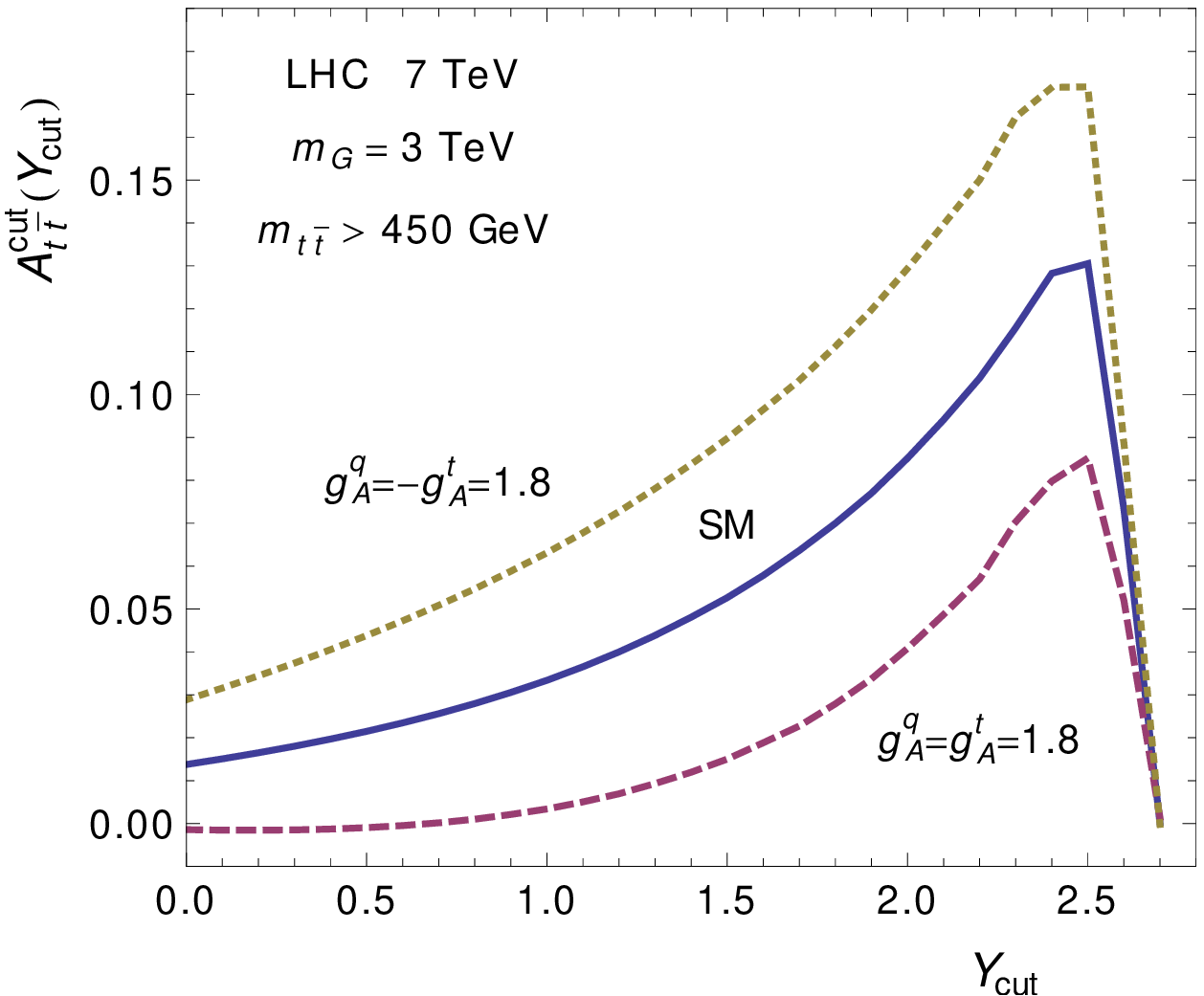}
\includegraphics[width=8cm]{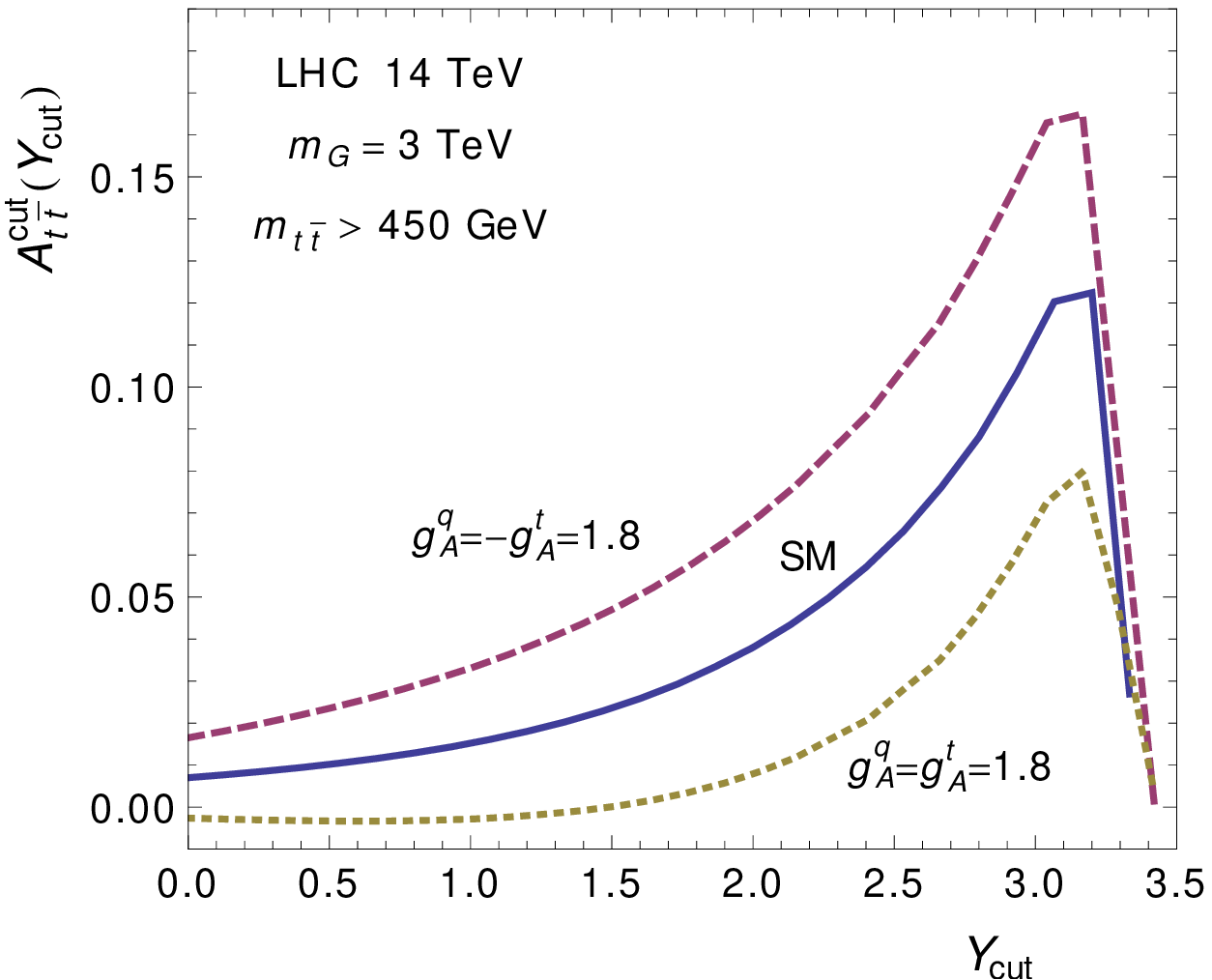}
\caption{Pair charge asymmetry $A_{t\bar t}$ at the LHC 
as a function of the mean rapidity $Y=(y_t+y_{\bar t})/2$, and
integrated pair charge asymmetry $A_{t\bar t}^{\rm cut}$
as a function of $Y_{\rm cut}$, with and without a cut in the invariant 
mass of the top quark pair $m_{t\bar t}$. 
Left plots with $\sqrt{s}=7$~TeV and right plots with $14$~TeV. 
\label{fig:paircutmA}}
\end{center}
\end{figure}

The pair charge asymmetries $A_{t\bar t}(Y)$ and $A_{t\bar t}(Y_{\rm cut})$ 
for both benchmark models and for $\sqrt{s}=7$~TeV and $14$~TeV 
are shown in Fig.~\ref{fig:paircutmA} in comparison with the SM prediction. 
We also provide predictions for samples with a large invariant mass of the 
top quark pair $m_{t\bar t} > 450$~GeV. As expected, the BSM contribution 
to the asymmetry in the octet U model is negative, and it is positive 
in the octet A model. For low values of $Y_{\rm cut}$ the integrated 
asymmetry $A_{t\bar t}(Y_{\rm cut})$ is almost twice the asymmetry in 
the SM for octet A, and almost vanishes for octet U. Introducing 
a cut in $m_{t\bar t}$ also enhances the size of the asymmetry, as in 
the SM, and in particular for large values of $Y_{\rm cut}$. 

\section{Summary and Conclusions}

The Standard Model predictions for the top quark charge asymmetry 
have been reanalysed including QED and weak corrections corrections.
For proton-antiproton collisions QED terms lead to an enhancement by 
a factor of about 1.2 in agreement with~\cite{Hollik:2007sw}, 
and slightly larger than the factor 1.1 obtained in~\cite{Kuhn:1998kw}. 
In total, our prediction is 
larger than the NLO Monte Carlo results by a factor around
$1.5\approx 1.2\times 1.3$, where the second factor arises from the 
different normalisation prescription.

The effect of a cut on the $t\bar t$ transverse momentum has been studied
and shown to lead to a significant enhancement of the asymmetry.
As a characteristic example we study a $p^{t\bar t}_\perp$-cut of 20~GeV 
which leads --- for the Tevatron --- to an enhancement of around 
1.3, and even more for more restrictive cuts.

Various definitions of observables are presented which are sensitive 
to the charge asymmetry and which can be measured at the LHC. 
The quantity $A_{t\bar t}(Y)$, which measures the forward--backward 
asymmetry with respect to the average rapidity of top and antitop quark, 
can amount up to $7\%$, if the large $Y$-region is selected. 
Considering the large statistics expected for the LHC in the near future, 
this asymmetry (and its integrated version) might soon be measurable 
at LHC experiments. We have also provided predictions in two benchmark 
axigluon-like models for these new observables.

\section*{Acknowledgements}

Work supported in part by the Research Executive Agency (REA) 
of the European Union under the Grant Agreement number 
PITN-GA-2010-264564 (LHCPhenoNet), 
by the Ministerio de Ciencia e Innovaci\'on under Grants
No. FPA2007-60323, FPA2011-23778 and PR2010-0481, 
by CPAN (Grant No. CSD2007-00042),
by the Generalitat Valenciana under Grant No. PROMETEO/2008/069, 
by the BMBF under contract 05HT4VKAI3, and 
the Sonderforschungsbereich/Transregio SFB/TR9 
''Computational Particle Physics''.
G.R. acknowledges hospitality at the Institut f\"ur Theoretische 
Teilchenphysik of the Karlsruher Institut f\"ur Technologie 
during the completion of this work. 
We thank J. Wagner-Kuhr, T. Peiffer, T. Chwalek, C. Boeser and J. Huston
for very usefull discussions. 


\end{document}